\documentstyle[12pt,aaspp4]{article}

\begin{document}

\newcommand{\gsim}{\mbox{\raisebox{-1.0ex}{$~\stackrel{\textstyle >}
{\textstyle \sim}~$ }}}
\newcommand{\lsim}{\mbox{\raisebox{-1.0ex}{$~\stackrel{\textstyle <}
{\textstyle \sim}~$ }}}
\newcommand{\psim}{\mbox{\raisebox{-1.0ex}{$~\stackrel{\textstyle \propto}
{\textstyle \sim}~$ }}}
\newcommand{\vect}[1]{\mbox{\boldmath${#1}$}}
\newcommand{\lmk}{\left(}
\newcommand{\rmk}{\right)}
\newcommand{\lnk}{\left\{ }
\newcommand{\nn}{\nonumber}
\newcommand{\rnk}{\right\} }
\newcommand{\lkk}{\left[}
\newcommand{\rkk}{\right]}
\newcommand{\lla}{\left\langle}
\newcommand{\p}{\partial}
\newcommand{\rra}{\right\rangle}
\newcommand{\vex}{{\vect x}}
\newcommand{\vek}{{\vect k}}
\newcommand{\vel}{{\vect l}}
\newcommand{\vem}{{\vect m}}
\newcommand{\ven}{{\vect n}}
\newcommand{\vep}{{\vect p}}
\newcommand{\veq}{{\vect q}}
\newcommand{\veX}{{\vect X}}
\newcommand{\veV}{{\vect V}}
\newcommand{\beq}{\begin{equation}}
\newcommand{\eeq}{\end{equation}}
\newcommand{\beqa}{\begin{eqnarray}}
\newcommand{\eeqa}{\end{eqnarray}}
\newcommand{\mpc}{\rm Mpc}
\newcommand{\kmpc}{Omega_0 h^2 {\rm Mpc}}

\title{Perturbative Analysis of Adaptive Smoothing Methods in
  Quantifying   Large-Scale Structure}
\author{\sc Naoki Seto }
\affil{Department of Physics, Faculty of Science, Kyoto University,
Kyoto 606-8502, Japan }

\begin{abstract}
Smoothing operation to make continuous density field from observed
point-like distribution of galaxies is crucially important for
topological or morphological analysis of the  large-scale structure, such
as, the genus statistics or the area statistics (equivalently the level
crossing statistics). 
It has been pointed out that the adaptive smoothing filters are more
efficient tools  to resolve cosmic
structures
  than the 
traditional spatially fixed filters. We study weakly nonlinear effects
caused by two representative  
adaptive methods often used in smoothed hydrodynamical particle (SPH)
simulations. Using framework of  second-order perturbation theory, we 
calculate the generalized skewness parameters for the adaptive methods
in the case of  initially power-law fluctuations.
 Then we apply the multidimensional Edgeworth expansion method
 and investigate weakly
 nonlinear evolution of the genus statistics and the area statistics.
Isodensity contour surfaces are often parameterized by the volume fraction
of the regions above a given density threshold.
We also discuss this parameterization method in perturbative manner.
\end{abstract}

\section{Introduction}
The large-scale distribution of galaxies is one of the most important
sources to study  formation and evolution of  cosmic structures. Now
there are two ongoing large-scale redshift surveys of 
galaxies, the Sloan Digital 
Sky Survey (SDSS, Gunn \& Weinberg 1995) and the Anglo-Australian
Telescope 2dF Survey (Colless 1998). These
surveys will bring us enormous information of three-dimensional galaxy
distribution and 
are expected to revolutionarily improve  our knowledge of
the large-scale structure in the universe.  

The two-point correlation function of galaxies, or its Fourier
transform, the 
power spectrum  are  simple as well as  powerful tools,  and have been 
widely used to quantify
 clustering of galaxies ({\it e.g.} 
Totsuji \& Kihara 1969, Peebles 1974, Davis \& Peebles 1983). These two
quantities 
are based on  the 
second-order moments of matter fluctuations. When the fluctuations are 
Random Gaussian distributed, the two-point correlation function or the power
spectrum contains full statistical information of the fluctuations.
   Even though the 
initial seed of structure
formation is often assumed to be random Gaussian distributed as
predicted by the standard inflation scenarios (Guth \& Pi 1982, Hawking
1982, Starobinsky 1982), this simple assumption has not been
observationally established. Moreover, the cosmic structures observed today
are more or less affected by nonlinear gravitational evolution. 
Therefore,    only with
 these two quantities, we cannot study the large-scale structure properly.

Other
statistical methods have  been proposed and is expected to play
complimentary roles to the 
traditional analyses based on the second-order moment.  The
skewness parameter characterizes  the asymmetry of one point
probability distribution function  (PDF) of the density  field and
has been investigated  in deep (Peebles 1980, Fry 1984, Juszkiewicz,
Bouchet \& Colombi 1993, Bernardeau 1994, Scoccimarro 1998, Seto
1999). Beside higher-order moments 
(such as skewness),
there are other statistical  
approaches designed to directly measure the   geometrical or 
morphological aspects of galaxy clustering.  Connectivity of the isodensity 
contour is an interesting target for these approaches.  For example,
 the  genus statistics were proposed by
Gott, Melott \& Dickinson (1986), the area statistics (equivalently the  level
crossing statistics) by Ryden (1988), and percolation analysis by Klypin 
(1988).   In addition, the Minkowski functionals recently attain much attention
({\it e.g.} Minkowski 1903, Mecke, Buchert \& Wagner  1994,  Schmalzing
\& Buchert 1997, Kerscher et al. 1997).

To  analyze observed cosmic structures, 
smoothing operation becomes crucially important in some cases. The
observed galaxies are 
distributed in point-like manner, but geometrical or
morphological analyses, such as, the genus statistics,  are usually  based on 
continuous (smoothed) density field (see also Babul \& Starkman 1992, 
Luo \& Vishniac 1995). We have traditionally used filters
with spatially fixed smoothing 
radius for analyzing  the large-scale
structure. 
Even though this method is  the simplest  from theoretical
point of views, other possibilities  are worth investigated.  
The local statistical fluctuations due to the discreteness of particles
are determined by the number of particles contained in 
 the smoothing kernel. If 
we use a spatially fixed filter, we can measure smoothed quantities at
overdensity regions relatively more accurately than  at underdense
regions.  As a result, quality of information becomes
inhomogeneous. This inhomogeneity is caused by the simple choice to
spatially fix the smoothing radius.
There must be more efficient methods to resolve cosmic 
structure from particle distribution (Hernquist \& Katz 1989).
Actually,
Springel et al. (1998) have pointed out that the signal to noise ratio of
the genus statistics is considerably improved by using adaptive smoothing
methods.
Adaptive methods are based on   Lagrangian description,  use nearly 
same number of ``particles" (mass elements) to construct smoothed
density field (Hernquist \& Katz 1989) and are expected to be less affected
by discreteness of mass 
elements. Therefore,  it seems reasonable that we can resolve cosmic structures
more efficiently, using these methods.

In this article, we perturbatively analyzed   quantitative effects
caused by  adaptive smoothing
methods. We pay  
special attention to three representative examples,
 the skewness parameter, the genus statistics and the area statistics.
 As the skewness is basically 
 defined by the  one point PDF, we can, in principle, 
 discuss it without making
 continuous density field. But its analysis is very  instructive to see 
 nonlinear effects accompanied with  adaptive smoothing methods.
 As a first step, 
we mainly study  the density field in  real space and 
do not discuss the  effects of biasing (Kaiser 1984, Bardeen, Bond, Kaiser
\& Szalay 1986, Dekel \& Lahav 1999). 

This article is organized as follows. In \S 2 we describe basic properties of
the adaptive smoothing and introduce its two main approaches. Then
perturbative formulas are derived for each of them.  In \S 3 we discuss 
the skewness parameters of the density field smoothed by these two 
adaptive methods.
We evaluate them using second-order 
perturbation theory. Some of  results in this section can be
straightforwardly
 applied
to the density field in the  redshift space.
Then we discuss weakly nonlinear effects of the  genus and  the area
statistics using the multidimensional Edgeworth expansion method explored  by
Matsubara (1994). To characterize the isodensity contour,
parameterization based on the volume fraction above a given density
threshold is often adopted. In \S 4.1 we discuss this parameterization in 
perturbative manner. In \S 4.2 we explicitly evaluate the generalized
skewness which is closely related to the genus and the  area statistics. In
\S 4.3 and \S 4.4, we
show the weakly nonlinear effects on these two statistics with various
smoothing methods. We make a brief summary in \S 5.

\section{Adaptive Smoothing Method}
The (unsmoothed) density contrast field $\delta(\vex)$  at a point
$\vex$  is defined in terms
of the mean density of the universe $\bar{\rho}$ and the local density
$\rho(\vex)$ 
as
\beq
\delta(\vex)=\frac{\rho(\vex)-\bar{\rho}}{\bar{\rho}}.
\eeq
In this article we assume that the primordial density fluctuations obey
Random Gaussian distribution which is completely characterized by the (linear) 
matter power 
spectrum.  Unless we state explicitly, we limit our analysis in the real 
space density field.  But some of our
results are straightforwardly applied to the redshift space quantities, 
as shown in the next section.

Isotropic   filters with  spatially constant smoothing radius $R$ have been
traditionally used to obtain continuous smoothed  density field
$\delta_{FR}(\vex) $ as follows
\beq
\delta_{FR}(\vex)=\int d\vex' \delta(\vex')W(|\vex'-\vex|;R).
\eeq
Here the function $W(\vex;R)$ is a spatial filter function and the
subscript $F$ indicate the  fixed smoothing. Most of  theoretical
analyses in the large-scale structure have been  based on this fixed smoothing 
method. As for the functional shape of $W(\vex;R)$,   two
kinds of  functions are often used ({\it e.g.} Bardeen et al. 1986,
Matsubara 1995).
 One is the Gaussian filter and defined as
\beq
W(\vex;R)=(2 \pi)^{-3/2}R^{-3} \exp\lmk -  {\vex^2}/{2R^2}\rmk.
\eeq
The other one is the top-hat filter and has a compact support as
\beqa \
W(\vex;R)=\cases{
3/(4\pi R^3) & $(|\vex|\le R)$ \cr
0 &  $(|\vex|> R).$ \cr
}\eeqa
In this article we mainly use the Gaussian filter. This filter is useful 
for quantifying the large-scale structure from observed noisy data sets.
In addition, algebraic
manipulations for the Gaussian filter are generally
 much simpler than for  the top-hat filter. 

Next let us discuss the   basic properties of   adaptive smoothing  methods 
(Hernquist \& Katz 1989, Thomas \& Couchman 1992, Springel et al. 1998).
The essence of these methods  is to change the smoothing radius
$R$ as a function of position $\vex$ according to its local  density contrast.
With a given spherically symmetric kernel $W$,  we determine the
smoothing radius 
$R(\vex)$ so that the  total 
mass included within  the kernel becomes  constant.
\beq
\bar{\rho}\int d\vex' (1+\delta(\vex'))R(\vex)^3
W(|\vex'-\vex|;R(\vex))=\bar{\rho}R^3. 
\eeq
The radius $R(\vex)$ becomes smaller than the standard value
$R$ in a overdense region and becomes
larger  in a underdense region.  In a system constituted by equal mass
particles as in standard  N-body simulations, the smoothing radius $R(\vex)$ is
determined so that the total number of particles in a filter becomes 
 constant. Thus  adaptive smoothing is   basically Lagrangian
 description and their smoothing radii 
are closely related to the resolution of spatial structures.

We can solve the variable smoothing radius 
 $R(\vex)$ in equation (5)  by perturbatively expanding 
the deviation  
$\delta R(\vex)\equiv R(\vex)-R$. In this procedure we  regard
the density contrast  $\delta$ as the order
parameter of the perturbative expansion. After some calculations we
 obtain the first-order solution 
as follows
\beq
\delta R(\vex)=-\frac13\delta_{FR}(\vex)R+O(\delta^2).
\eeq
This simple result seems quite  reasonable with the 
relation below. 
\beq
R(\vex)^3 (1+\delta_{FR}(\vex))=R^3 (1+O(\delta^2)).
\eeq
This  relation
  roughly shows  that the total mass within the smoothing radius 
$R(\vex)$ does not
depend on  position $\vex$.

With  the variable smoothing radius $R(\vex)$  (solution
for eq.[5])  we can
practice  adaptive smoothing. 
As pointed out by Hernquist \& Katz (1989) for  the smoothed
particle hydrodynamics (SPH), there exist two different methods 
({\it  gather} and {\it scatter} approaches) to
assign the  smoothed density contrast field
at  each point $\vex$. The gather approach is simply use the solution
$R(\vex)$ at the point $\vex$ in interest and the smoothed field is
 formally written as  
\beq
\delta_{GR}(\vex)=\int d\vex' \delta(\vex')W(|\vex'-\vex|;R(\vex))-C(R),
\eeq
the subscript $G$ indicates the gather approach. In this case,  the volume
average of the first term in the right hand side dose not vanish and  we
have added a term $C(R)$ so that the total volume 
average of $\delta_{GR}(\vex)$ becomes zero.

In the scatter approach  we use the solution $R(\vex')$ for each point where 
a mass element exists. We can write down the smoothed field at $\vex$ as
\beq
\delta_{SR}(\vex)=\int d\vex' \delta(\vex')W(|\vex'-\vex|;R(\vex')),
\eeq
the subscript $S$ represents the scatter approach. In this case the
volume average becomes zero. We  only dilute the mass element at point
$\vex'$ with the density profile proportional to 
$W(|\vex'-\vex|;R(\vex'))$ around that
point. 
Note that the spatial dependence of the smoothing radius $R(\cdot)$ is
different between equations (8) and (9).

Next  we  evaluate 
 equations for $\delta_{GR}(\vex)$ and $\delta_{SR}(\vex)$  up to 
second-order of the  density contrast 
$\delta$ using   perturbative solution of the smoothing radius 
 $R(\vex)=R+\delta R(\vex)$
given in equation (6). The results are given as  
\beqa
\delta_{GR}(\vex)&=&\delta_{FR}(\vex)-\frac13\delta_{FR}(\vex)R\frac{\p}{\p 
  R}\delta_{FR}(\vex)+\frac16 R\frac{d}{d 
  R}\sigma_R^2+O(\delta^3),\\
\delta_{SR}(\vex)&=&\delta_{FR}(\vex)-\frac{R}3\int  d\vex' \p_R
W(|\vex'-\vex|;R)\delta(\vex')\delta_{FR}(\vex') +O(\delta^3).
\eeqa
The formula for the scatter approach is somewhat complicated, compared
with the gather approach. Also in numerical analysis, the  scatter approach 
requires  higher computational costs (Springel et al. 1998).  This  reflects
 nonlocal character of the smoothing radius.

Equations (10) and (11) show apparently  that 
 the corrections due to the adaptive methods start from 
second-order of $\delta$. Therefore,  their effects are expected to be
 comparable to 
second-order (nonlinear) effects predicted by  cosmological gravitational
perturbation theory (Peebles 1980, Fry 1984,  Goroff et al. 1986).
 Adaptive smoothing methods  modify the   quantities 
 which  characterize  the 
nonlinear mode couplings, such as the  skewness parameter of  density field.

If we use  the Gaussian filter,  the leading-order correction  for the 
 scatter approach is expressed 
 as follows
\beq
\int\frac{d\vek}{(2\pi)^3}\frac{d\vel}{(2\pi)^3}
\exp\lkk-\frac{(2\vel^2+\vek^2+2\vek\cdot \vel)R^2}{2} \rkk \delta(\vek)
\delta(\vel)\exp[i(\vek+\vel)\cdot \vex]\frac{(\vek+\vel)^2R^2}{3},  
\eeq
where $\delta(\vek)$ and $\delta(\vel)$ are the Fourier coefficients of
 the density contrast and defined  as  
\beq
\delta(\vek)=\int d\vex\delta(\vex)
\exp(-i\vek\cdot\vex).
\eeq
Formula (12) is useful to quantitatively evaluate  nonlinear effects
 caused by the scatter approach.

\section{Skewness}
In this section we investigate  modifications of the skewness
parameter $S$ caused by the two adaptive methods.  Skewness is a fundamental
quantities to 
characterize  asymmetry of the one point PDF of the  density 
field (Peebles 1980, Fry 1984, Juszkiewicz,
Bouchet \& Colombi 1993, Bernardeau 1994).  It is defined as
\beq
S=\frac{\lla \delta^3\rra}{\sigma^4},
\eeq
where the angular  bracket $\lla \cdot\rra$ represents to take the ensemble
average and $\sigma(\equiv\lla \delta^2\rra^{1/2})$ is the rms
fluctuation of $\delta$.  Here,  we 
discuss the leading-order contributions for  the numerator $\lla
\delta^3\rra$ and denominator $\sigma^4$.
As we have already commented in \S 1, the skewness
parameter can be discussed 
 without making continuous density field. It can be basically
defined by the count probability distribution function,  and  spatial
relation between one region and another one is unnecessary ({\it e.g.}
Gazta$\tilde{\rm n}$aga 1992, Bouchet et al. 1993, Kim \& Strauss 1998, and
references therein, see also Colombi, Szapudi \& Szalay 1998).
 Therefore our 
effort in this article to resolve cosmic structures  by using the  adaptive
smoothing might be irrelevant for  observational determination of
the skewness parameter.  But perturbative analysis in this section is
very useful to 
grasp nonlinear effects caused by the adaptive smoothing methods and
become basis for
studying statistics of isodensity contours  such as the genus statistics or
the area statistics discussed in the next section.

The leading-order contribution for  the rms fluctuation 
 $\sigma$ is written  in terms of the linear (primordial) power
spectrum $\lla \delta_1(\vek) \delta_1(\vel)\rra=(2\pi)^3\delta_{Drc}(\vek+\vel)P(k)$ ($ \delta_1(\vek)$ : linear
mode, $\delta_{Drc}(\cdot)$: Dirac's delta function). With the Fourier transformed filter function $w(kR)$  we have
\beq
\sigma_R^2=\lla \delta^2_R(\vex)\rra=\int
\frac{d\vek}{(2\pi)^3}P(k)w(kR)^2,
\eeq
where  the suffix $R$ is added  to explicitly indicate the
smoothing radius $R$.
Throughout in this article,  we use  power-law spectra  $P(k)$ as
\beq
P(k)=Ak^n,~~~-3<k\le 1.
\eeq
for these scale-free models the normalization factor $A$ becomes
 irrelevant and we can 
simply put $A=1$ below. Furthermore, as shown later, 
 the skewness parameter does not
 depend on the smoothing radius in our leading-order analysis.
 From equation (15)  we have the variance $\sigma_R^2$ for the Gaussian
filter as
\beq
\sigma_R^2=\int_0^{\infty}\frac{dk}{2\pi^2}k^{n+2}e^{-k^2 R^2}=\frac{R^{-n-3}}{(2\pi)^{2}}\Gamma\lmk\frac{3+n}{2} \rmk. 
\eeq
The integral  (15) logarithmically diverges for $n=-3$, but skewness  $S$
 is well-behaved  in the limit
 $n\to -3$ from above. As it shows interesting behavior  at this specific
 spectral index,  we also discuss quantities at  $n=-3$ regarding them
 as the limit values.

Calculation of the third-order moment $\lla\delta^3\rra$ is more
complicated than that of the variance  $\sigma^2$ discussed so far.
 When the initial fluctuation is random Gaussian distributed as
assumed in this article, the 
linear contribution for the third-order moment 
becomes exactly zero due to the symmetric
distribution of the density contrast $\delta$ around the origin
$\delta=0$. Nonlinear mode 
couplings induce asymmetry in this distribution.
Therefore, we resort to   higher-order perturbation theory. The
leading-order contribution for the 
skewness parameter without smoothing operation  is
given by Peebles (1980) in the  case of Einstein de-Sitter background   as
\beq
S=\frac{34}7.
\eeq
It is convenient to use the  Fourier space  representation to calculate
the third-order moment 
 for the smoothed density field.   Following the standard procedure, 
we expand a nonlinear Fourier modes  of overdensity $\delta$ and the 
(irrotational) peculiar velocity field $\veV$ as (Fry 1984, Goroff et al. 1986)
\beqa
\delta(\vex)&=&\delta_1(\vex)+\delta_2(\vex)+\cdots,\nonumber\\
\veV(\vex)&=&\veV_1(\vex)+\veV_2(\vex)+\cdots,
\eeqa
where $\delta_1(\vex)$ and $\veV_1(\vex)$ are  the linear modes and
$\delta_2(\vex)$ and $\veV_2(\vex)$ the
second-order modes.
We perturbatively solve the  continuity, Euler and Poisson
equations,
\beqa
\frac{\p}{\p
  t}\delta(\vex)+\frac{1}{a}\nabla[\veV(\vex)\{1+\delta(\vex)\}]&=&0,\nonumber\\ 
\frac{\p}{\p
  t}\veV(\vex)+\frac1a[\veV(\vex)\cdot\nabla]\veV(\vex)+\frac{\p_t
  a}a\veV(\vex)+\frac1a\nabla\phi(\vex)&=&0,\nonumber\\
\nabla^2\phi(\vex)-4\pi a^2\rho(t)\delta(\vex)&=&0,\nonumber
\eeqa
where $a$ represents the scale factor.
The second-order solution in $\vek$-space is given as 
\beq
\delta_2(\vek)=\int\frac{d\vel}{(2\pi)^3}\delta_1(\vel)\delta_1(\vek-\vel) 
J(\vel,\vek-\vel),
\eeq
or in  $\vex$-space
\beq
\delta_2(\vex)=\int\frac{d\vek}{(2\pi)^3}\frac{d\vel}{(2\pi)^3}e^{i\vek\cdot 
  \vex}\delta_1(\vel)\delta_1(\vek-\vel) 
J(\vel,\vek-\vel),
\eeq
where the kernel $J$ is defined by 
\beq
J(\vek,\vel)=
\frac12(1+K)+\frac{\vek\cdot\vel}2\lmk\frac1{k^2}+\frac1{l^2}\rmk
+\frac12(1-K) \frac{(\vek\cdot\vel)^2}{k^2l^2}.
\eeq
The factor $K(\Omega,\lambda)$  weakly depends on the
density parameter $\Omega$ and cosmological constant $\lambda$ as  shown 
in the fitting formula (Matsubara 1995, see also Bouchet et al. 1992)
\beq
K(\Omega,\lambda)\simeq\frac37\Omega^{-1/30}-\frac{\lambda}{80}\lmk1
-\frac32\lambda\log_{10}\Omega\rmk.  
\eeq
In the ranges of two parameters $\Omega$ and $\lambda$
\beq
0.1\le \Omega\le1,~~~0.1\le \lambda\le1,
\eeq
 the
difference of $K(\Omega,\lambda)$ from 
$K=3/7$ is within
$8\%$. Therefore, in the following analysis  we basically study the 
Einstein de-Sitter background and  use $K=3/7$.

Using the second-order solution (21) we can derive the well known formula for the
third-order moment as follows (Juszkiewicz et al. 1993) 
\beq
\lla\delta_R^3\rra=6\int\frac{d\vek}{(2\pi)^3}\frac{d\vel}{(2\pi)^3}
P(k) P(l)J(\vek,\vel)w(kR)w(lR) w(|\vek+\vel|R).
\eeq
Let us simplify this six-dimensional integral $d\vek d\vel$. In the case of the
Gaussian filter 
\beq
w(kR)=\exp(-k^2R^2/2),
\eeq
we can change $\lla\delta_R^3\rra $ to the following form (Matsubara 1994)
\beq
\lla\delta_R^3\rra=\frac3{28\pi^4}(5I_{220}+7I_{131}+2I_{222}),
\eeq
where we have defined
\beq
I_{abc}=\int_0^\infty dk \int_0^\infty dl\int_{-1}^1
du\exp[-R^2(k^2+l^2+ukl)]k^al^bu^c P(k) P(l).
\eeq
For a power-law initial fluctuation $P(k)=k^n$,  we  obtain a  final closed
formula
(Matsubara 1994, $\L$okas et al.  1995)
\beq
S_F(n)=3F\lmk \frac{n+3}2,
\frac{n+3}2,\frac32;\frac14\rmk-\lmk n+\frac87\rmk F\lmk \frac{n+3}2,
\frac{n+3}2,\frac52;\frac14\rmk,\label{a26}
\eeq
where $F$ is the Hypergeometric function.

In the case of the top-hat filter whose Fourier transform is given by 
\beq
w(kR)=\frac3{(kR)^3}(\sin kR-kR\cos kR),
\eeq
the final form of $S$ becomes very simple as follows (Juszkiewicz et al. 1993,
Bernardeau 1994)
\beq
S_F(n)=\frac{34}7-(n+3).
\eeq
This formula is not only valid for pure  power-law initial fluctuations but
also for general power spectra with effective spectral index defined at 
the smoothing radius $R$ as
\beq
n\equiv-\frac{d\ln \sigma_R^2}{d\ln R}-3.
\eeq

Equations (29) and (31) are only the leading-order contribution and more
higher-order effects might 
change them considerably. Thus it is quite important 
to compare these analytic formulas with fully  nonlinear numerical
simulations and 
clarify  validity of the perturbative formulas.
 There are many works on this topic and the
analytic  predictions  show surprisingly  good agreement with numerical
simulations,  even at $\sigma\sim1$ ({\it e.g.} Baugh,
Gazta$\tilde{\rm n}$aga \& Efstathiou 1995, Hivon et al. 1995,
Juszkiewicz et al. 1995, $\L$okas et al. 1995).

So far we have discussed skewness $S$ with fixed smoothing methods. For the
third-order moments $\lla\delta_R^3\rra$, the second-order effects
caused by the 
gravitational evolution and that caused by the adaptive smoothing are
decoupled,  as we can see from equations (10) and (11). Thus we can write the
skewness parameter for  adaptive methods  in the following forms
\beqa
S_{G}&=&S_{F}+\Delta S_{G},\\
S_{S}&=&S_{F}+\Delta S_{S}.
\eeqa
Here $\Delta S_{G}$ and $\Delta S_{S}$ are the correction terms caused
by 
the  adaptive smoothing methods. In the next two subsections we
calculated these 
terms explicitly.

\subsection{Gather Approach}
First we calculate the correction term $\Delta S_{G}$ for the gather
approach. With equation (10) this term is easily transformed to the
 following equation (see Appendix A.1 
for 
derivation)
\beq
\Delta S_G=-\frac{d\ln \sigma_R^2}{d\ln R}.
\eeq
For a power-law spectrum we have simple equation below 
\beq
\Delta  S_{G}(n)=(n+3).
\eeq
In derivation of  equation (35)
  we only use  Gaussianity of the  one point PDF of
the linear smoothed field $\delta_R$. Therefore, these formulas
 do not depend on the shape of the smoothing
filter nor the cosmological parameters $\Omega$ or
$\lambda$. Furthermore they are valid also 
 in the  redshift space, if we use the  the distant observer
approximation.  Thus equation (35) has strong predictability.

Hivon et al. (1995) perturbatively  examined the skewness parameters in
redshift space 
and evaluate them both for the top-hat filter and the 
 Gaussian filter. They also compared their analytic results with
 numerical results.  They  found  that
 these two show agreement only in the range $\sigma\lsim 0.1$, in
 contrast  to the  skewness parameter in the real space 
 $\sigma\lsim1.0$.  They commented that this limitation is mainly due to the
 finger of god effects ({\it e.g.} Davis \& Peebles 1983).

Here we use their analytic results
 and combine our new formula with them. In figure 1 we
present the skewness parameters for various spectral indexes $n$ both in
 the real
and redshift spaces.  For simplicities we limit our analysis for the Einstein
 de-Sitter background.

For the Gaussian filter, the  skewness parameter 
 by the gather method is a increasing
function of spectral index $n$ both in real and redshift spaces. This
dependence is contrast to the skewness with
the fixed  smoothing method.  
Comparing the skewness in the real and the  redshift spaces, 
$n$ dependence of the gather method is 
 somewhat weaker in real space  than in  redshift space, but this tendency is
 also different 
 from  the fixed smoothing.

For the top-hat filter, there is no spectral index  $n$- dependence  in
  the real space.
  We have $S=34/7$ which is the same as the unsmoothed value (Peebles
  1980). Bernardeau (1994) pointed out that the  skewness $S$  filtered with
  the  top-hat filter in
  Lagrangian space does not depend on the power spectrum and is given by
  $S=34/7$.
As the adaptive smoothing is basically Lagrangian description, this fact
seems reasonable.   
In the case of  the 
redshift space we have a  fitting formula below
\beq
S_G(n)=\frac{35.2}7-0.15(n+3),~~~({\rm Einstein~de~Sitter~background})
\eeq
which is based on  formula (49) of  Hivon et al (1995).
Again $n$ dependence is very weak and
 becomes weaker for $\Omega<1$ (see Fig.4 of Hivon et
al. 1995). 
  Finally,  we comment the possibility that our
perturbative treatment of  the redshift space skewness   becomes
worse  in the adaptive methods than in the  fixed smoothing method. In the
adaptive methods, smoothing radius of a high density region becomes
smaller and the (strongly nonlinear) finger of god effects might not be
suppressed well.

\subsection{Scatter Approach}
Next we calculate the correction term $\Delta S_S$ for the scatter
approach. We only discuss the real space density field  smoothed with the
Gaussian filter (eq.[3]). From
equation (12) we obtain the following equation (see Appendix A.1),
\beq
\Delta  S_{SR}(n)\sigma_R^4=2\int\frac{d\vek}{(2\pi)^3}\frac{d\vel}{(2\pi)^3}
\exp\lkk-\frac{(3\vel^2+2\vek^2+2\vek\cdot \vel)R^2}{2} \rkk
P(k)P(l){(\vek+\vel)^2} R^2.
\eeq 
The six dimensional integral $d\vek d\vel$ is  simplified to a
three dimensional integral $dkdldu$ as in equations (25) and (28). Then we
have the  following relation
\beqa
\Delta  S_{SR}(n)\sigma_R^4&=&\frac1{4\pi^4}\int_0^\infty
{dk}\int_0^\infty{dl}\int_{-1}^1du 
\exp\lkk-\frac{(3l^2+2k^2+2klu)R^2}{2} \rkk \nn\\
& &\times k^2 l^2
P(k)P(l){(k^2+l^2+2klu)} R^2.
\eeqa
For  a  pure-power law fluctuation we obtain the following
analytic formula
\beqa
\Delta  S_{S}(n)&=&-2^{(n+3)/2}3^{-(n+5)/2} (n+3)  \bigg\{
 \frac23(n+3)F\lmk\frac{5+n}{2}, \frac{5+n}{2},\frac{5}{2},\frac{1}{6} \rmk \nn\\
& & - {5}
F\lmk \frac{3+n}{2}, \frac{5+n}{2},\frac{3}{2},\frac{1}{6}\rmk  \bigg\}.
\eeqa
\if0%%%%%%%%%%%%%%%%%%%%%%%%%%%%%%%%%%%%%
\beqa
\Delta  S_{S}(n)&=&-2^{(n+3)/2}3^{-(n+5)/2}  \bigg\{ 12F\lmk\frac{3+n}{2},
  \frac{3+n}{2},\frac{1}{2},\frac{1}{6}\rmk\nn\\ & & - 12F\lmk\frac{3+n}{2},
  \frac{3+n}{2},\frac{3}{2},\frac{1}{6} \rmk
- {5(3+n)} 
F\lmk \frac{3+n}{2}, \frac{5+n}{2},\frac{3}{2},\frac{1}{6}\rmk \bigg\}
\eeqa
\fi%%%%%%%%%%%%%%%%%%%%%%%%%%%%%%%%%%%%%%%
In contrast  to the previous gather approach, this result is valid only to
  the real space skewness   with  the Gaussian filter. In table 1 we
 present  numerical values of  $\Delta  S_{S}(n)$.  In figure 2 we show
  $ S_{S}(n)$ as a function of the spectral index $n$.
We can see that 
$n$ dependence is similar to the gather approach but now it becomes  weaker.
If we change $n$ from $-3$ to $-1$, skewness $S$ changes $\sim 25\%$ for the 
 scatter approach, $\sim 45\%$ for the gather approach, and $\sim 38\%$
 for the fixed smoothing.

\section{Statistics of Isodensity Contour}
The genus number  is a topological quantity and  defined
by the number of the homotopy classes of closed curves that may be drawn 
on a surface without cutting them into two pieces.  This definition
seems highly mathematical, but there are more intuitive methods to count 
the 
genus number. First one is to notice the number of holes and
isolated regions of the surface in interest. Second one is to count 
stationary points of the surface along one spatial direction (Adler
1981, Bardeen et al. 1986).
With these equivalent methods, we can calculate the genus density as
follows
\beqa
{\rm Genus~ density}&=&\frac{N({\rm holes})-N({\rm isolated
    ~regions})}{\rm volume}\\
 &=&-\frac{N({\rm maxima})+N({\rm minima})-N({\rm saddle~
     points})}{\rm2\times  volume}.
\eeqa
 For example, in the case
of  one-sphere, we have $N({\rm holes})=0$ and $N({\rm isolated
    ~regions})=1$ and genus number becomes $-1$. We obtain the same result 
  with equation (42). 
The genus number density of isodensity contour of the  large-scale
structure is a powerful measure to quantify   connectivity of galaxy
clustering, such as, filamentary networks, sheet-like or bubble-like
structures.   The genus density of a high density contours
is expected to be
 negative as the surfaces would 
show disconnected meatball-like structure. But the  genus density for
contours around  the  mean density $\delta\sim0$ would be positive as
they would 
look like highly connected sponge-like structure (Gott, Melott \&
Dickinson 1986). 
The genus density as a function of the  matter density threshold is called
the genus  
statistics and has been widely investigated both numerically and
observationally 
(Gott, Weinberg \& Melott 1987, Weinberg, Gott \& Melott 1987, Melott,
Weinberg \& Gott 1988, Gott et al. 1989, Park \& Gott 1991, Park, Gott
\& da Costa 1992, Weinberg \& Cole 1992, Moore et al. 1992, Vogeley,
Park, Geller, Huchra \& Gott 1994, Rhoads, Gott \& Postman 1994,
Matsubara \& Suto 1996, Coles, Davies \& Pearson 1996, Sahni et
al. 1997, 
Protogeros \& Weinberg 1997, Coles, Pearson, Borgani, Plionis \&
Moscardini 1998, Canavezes et
al. 1998, Springel et al. 1998)

Usually we use the local expression (42) to analytically study the genus
statistics. For the genus density of isodensity contour at
$\nu\equiv \delta/\sigma$,  this expression  is written as follows
(Doroshkevich 1970, Adler 1981, Bardeen et al. 1986, Hamilton, Gott \&
Weinberg 1986)
\beq
G(\nu)=-\frac12\lla\delta_{Drc}[\delta(\vex)-\nu\sigma]\delta_{Drc}[\p_1\delta(\vex)]\delta_{Drc}[\p_2\delta(\vex)]|\p_3\delta(\vex)|(\p_{11}\delta(\vex)\p_{22}\delta(\vex)- 
\p_{12}\delta(\vex)^2 
) \rra,
\eeq 
where $\delta_{Drc}(\cdot)$ represents the Dirac's delta function. The
first one $\delta_{Drc}[\delta(\vex)-\nu\sigma]$ specifies the contour
$\nu\equiv \delta/\sigma$. The second and third ones
$\delta_{Drc}[\p_1\delta(\vex)]$, $\delta_{Drc}[\p_2\delta(\vex)]$
specify the stationary points along $x_3$ direction. The term 
$(\p_{11}\delta(\vex)\p_{22}\delta(\vex)- 
\p_{12}\delta(\vex)^2 
)$ is the determinant of the Hesse-matrix and assigns  proper
signatures for  the stationary points corresponding to  signs of equation 
(42). Even though equation (43) introduces a specific spatial direction
($x_3$-axis), Seto et al. (1997) derived a
rotationarilly symmetric formula, and studied nonlinear evolution of
the genus statistics   using the  Zeldovich approximation (Zeldovich 1970)

In the case of  an isotropic random Gaussian fluctuation which is usually
assumed as the initial condition of the structure formation, the
complicated formula (43) is simplified to (Doroshkevich 1970, Adler 1981,
Bardeen et al. 1986, Hamilton, Gott \& 
Weinberg 1986)
\beq
G(\nu)=\frac1{(2\pi)^2}\lmk\frac{\sigma_1^2}{3
  \sigma^2}\rmk^{3/2} e^{-\nu^2/2}(1-\nu^2),
\eeq
where $\sigma_1^2$ is defined as
\beq
\sigma_1^2=\lla(\nabla\delta)^2 \rra=\frac1{2\pi^2}\int_0^\infty dk
P(k)k^4 w(kR)^2.
\eeq
Nonlinear evolution of the genus statistics had been studied using N-body
simulations, but most analytical predictions for the genus statistics have been based on
the linear formula (44). To compare it with observed distribution of
galaxies we have to use sufficiently large smoothing radius to reduce
nonlinearities. However, such a large smoothing radius is not
statistically preferable for the finiteness of our survey volume. 

Matsubara (1994) improved this difficulty by taking into account of  weakly 
nonlinear 
effects in the genus statistics (see also Hamilton 1988, Okun 1990, 
Matsubara \& Yokoyama 1996,
Seto et al. 1997). He used the  multidimensional Edgeworth
expansion method and added the first-order nonlinear correction to
the linear formula.  His result is written as  
\beq
G(\nu)=-\frac1{(2\pi)^2}\lmk \frac{\sigma_1^2}{3
  \sigma^2}\rmk^{3/2}e^{-\nu^2/2}\lkk H_2(\nu)+\sigma \lmk\frac{S}6
H_5(\nu) +\frac{3T}{2} H_3(\nu) +3UH_1(\nu) \rmk+O(\sigma^2)\rkk.
\eeq
This formula is valid for  statistically isotropic and homogeneous
weakly random Gaussian fields.
Here functions 
$H_n(\nu)\equiv (-1)^ne^{\nu^2/2}(d/d\nu)^n e^{-\nu^2/2}$
are  the Hermite polynomials,
\beqa
H_0(x)&=&1,~~H_1(x)=x,~~~H_2(x)=x^2-1,\nn \\
H_3(x)&=&x^3-3x,~~H_5(x)=x^5-10x^3+15x.
\eeqa
In equation (46), $S$ is the skewness parameter discussed in the previous
section.  $T$ and $U$ are called  the generalized skewness parameters  and 
defined by
\beqa
T&=&-\frac1{2\sigma^2\sigma_1^2}\lla\delta^2\Delta\delta\rra, \nn \\
U&=&-\frac3{4\sigma_1^4}\lla\nabla\delta\cdot\nabla\delta\Delta\delta\rra. 
\eeqa
Matsubara \& Suto (1996) examined the perturbative formula (46) using
N-body simulations. For power-law spectra with $n=1,0$ and $-1$, they found
that this formula are in reasonable agreement with numerical results
 in the range $-0.2\lsim
\nu\sigma\lsim 0.4$. 

Next let us briefly summarize the area statistics $N_3(\nu)$ which were
proposed by Ryden (1988) and investigated  detailedly by Ryden et 
al. (1989).
The area statistics are defined as the mean area of isodensity contour
surface per unit volume. For statistically homogeneous and isotropic
fluctuations, the area statistics are equal to twice the mean number of
isodensity contour crossings along a straight line of unit length. 
These two statistics are thus equivalent (beside factor 2), but the
contour crossing statistics are easier to compute numerically (Ryden 1988). 

As in equation (43), the area statistics for isodensity contour
$\delta=\nu\sigma$ is 
written as
\[
N_3(\nu)=\lla \delta_{Drc}[\delta(\vex)-\nu\sigma]|\nabla\delta(\vex)|\rra.
\]
In the case of isotropic Random Gaussian fluctuations, we have the  following 
formula (Ryden 1988)
\beq
N_3(\nu)=\frac2\pi \lmk\frac{\sigma_1^2}{3\sigma^2} \rmk^{1/2}e^{-\nu^2/2}.
\eeq

Weakly nonlinear effects on the area statistics can be  discussed 
 with a similar technique used to derive equation (46) (Matsubara
 1995). In this  
case, we need information of the density field up to its   first spatial
derivative  and nonlinear correction is expressed in terms of
 two parameters $S$ and $T$ as follows
\beq
N_3(\nu)=\frac2\pi \lmk\frac{\sigma_1^2}{3\sigma^2}
\rmk^{1/2}e^{-\nu^2/2}
\lkk 1+\sigma\lmk\frac{S}6 H_3(\nu)+\frac{T}2H_1(\nu) \rmk +O(\sigma^2)\rkk.
\eeq

In the rest of this section we consider nonlinear  effects of the
adaptive smoothing methods  on the 
genus and the area statistics. We calculate the generalized skewness parameters $T$
and $U$ both for the gather approach and the scatter approach.
We limit our analysis to the   real space density field smoothed by
Gaussian filters.

\subsection{Reparameterization of Isodensity Contour}

Equation (46) is weakly non-Gaussian genus density for isodensity
contour surfaces
parameterized by the  simple definition $\nu=\delta/\sigma$. There is another 
conventional 
method to name  contour surfaces. In this method,  we notice the volume
fraction  
$f$ above the   density threshold of the contour in interest ({\it e.g.}
Gott, Melott \& Dickinson 1986, Gott et al. 1989), and 
parameterize  the
contour using   value $\nu_r$ defined by  
\beq
\nu_r\equiv {\rm erf}^{-1}(f),
\eeq
where the suffix $r$ indicates `` reparameterization''  and ${\rm erf}(x)$
is the error 
function defined by
\beq
{\rm erf}(x)\equiv \frac1{\sqrt{2\pi}}\int_x^\infty dy e^{-y^2/2}.
\eeq
This procedure is a kind of Gaussianization. Two methods  coincide
$\nu=\nu_r$ when the one point PDF
$P(\nu)$ is Gaussian distributed. If we use  this new parameterization,
the 
genus curve is 
apparently 
invariant under a monotonic mapping of the density contrast field
$\delta$. Furthermore,  it has been long known that the  genus curve
with  $\nu_r$
parameterization
(51) nearly  keeps its original symmetric shape (eq.[44])  in the course of
 weakly nonlinear gravitational
evolution 
of  density field ({\it e.g.} Springel et al. 1998 and references
therein).  Almost the  same 
kind of tendency has been confirmed for the area statistics (Ryden et
al. 1989). Weakly nonlinear area density  with $\nu_r$ parameterization 
 remains at its linear  shape very well. Here,  let us
relate these two parameterization methods for 
 weakly nonlinear regime.  Using the Edgeworth expansion method, the one
point PDF $P(\nu)$ is written in terms of the skewness $S$   up to the
first-order 
nonlinear correction 
\beq
P(\nu)=\frac1{\sqrt{2\pi}}e^{-\nu^2/2}\lkk
1+\frac{\sigma S}6H_3(\nu)+O(\sigma^2)\rkk. 
\eeq
Juszkiewicz et al. (1995) examined this approximation using N-body
simulations. They found that the  above formula is accurate until
$\sigma S$ reaches 1. Therefore the inequality
\beq
\sigma\lsim S^{-1}\sim0.2,
\eeq
 would be a standard for the validity of the
perturbative analysis in this subsection.
The volume fraction $f(\nu)$ above the threshold $\delta=\nu
\sigma$ is given by
\beq
f(\nu)=\int_\nu^\infty dx P(x)=
{\rm erf}(\nu)+\frac{\sigma S}{6\sqrt{2\pi}}\lnk e^{-\nu^2}(\nu^2-1)
\rnk+O(\sigma^2).
\eeq
With equations (51) and (55) we  obtain  correspondence
between $\nu$ and $\nu_r$ as follows
\[
\nu=\nu_r+\frac{\sigma S}6 \lnk\nu_r^2-1 \rnk+O(\sigma^2).
\]
Finally the genus density $G_r(\nu_r)$ in this new parameterization is
given by $G(\nu)=G_r(\nu_r)$ and written as
\beqa
G_r(\nu_r)&=&-\frac1{(2\pi)^2}\lmk \frac{\sigma_1^2}{3
  \sigma^2}\rmk^{3/2}e^{-\nu_r^2/2}\bigg[ H_2(\nu_r)+\sigma \bigg(
H_3(\nu)\lmk-S+\frac32 T\rmk  \nn\\
& & +H_1(\nu)(-S+3U) \bigg)+O(\sigma^2)\bigg].
\eeqa

Similarly  we have the following result for the area statistics
\beq
N_{3r}(\nu_r)=\frac2\pi \lmk\frac{\sigma_1^2}{3\sigma^2}
\rmk^{1/2}e^{-\nu_r^2/2}
\lkk 1+\sigma\lmk-\frac{S}3 +\frac{T}2 \rmk H_1(\nu_r)+O(\sigma^2) \rkk.
\eeq
In this case, the first nonlinear correction is simply proportional to
$\nu_r$ (see eq.[47]) and it completely vanishes when we have $S=3/2T$.

Later  in \S 4.3 and \S4.4, we will confirm that the nonlinear correction
(proportional to $\sigma$) for the  genus and area statistics with the 
 fixed smoothing method are very small for   $\nu_r$ parameterization,
as experientally known  in N-body simulations. 
In the followings,  we use these two parameterizations $\nu$ and $\nu_r$.

\subsection{Generalized Skewness}
The generalized skewness $T$ and $U$ are basic ingredients to
perturbatively evaluate the weakly non-Gaussian effects on the
 genus and the  area statistics. For the  Gaussian
filter with a fixed smoothing radius,  explicit formulas
valid  for the   power-law initial
fluctuations
 were derived by Matsubara
(1994).
They are given as follows
\beqa
T_{FR}&=&
3F\lmk \frac{n+3}2,\frac{n+5}2,\frac3 2,\frac14 \rmk
-\lmk n+\frac{18}7\rmk F\lmk \frac{n+3}2,\frac{n+5}2,\frac5 2,\frac14
\rmk\nn\\
& &+\frac{4(n-2)}{105}F\lmk \frac{n+3}2,\frac{n+5}2,\frac7 2,\frac14
\rmk,\\
U_{FR}&=&F\lmk \frac{n+5}2,\frac{n+5}2,\frac5 2,\frac14\rmk
-\frac{7n+16}{35} F\lmk \frac{n+5}2,\frac{n+5}2,\frac7 2,\frac14
\rmk.
\eeqa
As shown in the case of  
 the  skewness parameter $S$ analyzed in \S 3.1 and \S 3.2, the 
second-order (first nonlinear)  effects caused by the adaptive smoothing 
methods are decoupled from those induced by gravitational mode
couplings. Thus we can calculate them separately and express the total
values in forms  similar to equations (33) and (34). 

First, we  analyze correction terms $\Delta T_{GR}$
and $\Delta U_{GR}$  for the gather approach. We  define these terms
by the  following equations
\beq
 T_{GR}= T_{FR}+\Delta T_{GR},~~ U_{GR}= U_{FR}+\Delta U_{GR}.
\eeq
 After some tedious algebra using  equation (10), we obtain the
 leading-order correction terms as follows  (Appendix A.2)
\beqa
\Delta T_{GR}&=&\frac23(n+4),\\
\Delta U_{GR}&=&\frac13(n+5).
\eeqa
These results are  similar to  the correction term for the 
skewness  $\Delta S_{GR}=n+3$ (eq.[36]) which
does not depend on the shape of the filter function. However, 
situation is not so much 
simple here. For the Gaussian filter we have the following relation, 
\beq
R\frac{\p \delta_R(\vex)}{\p R}=R^2\Delta_{x} \delta_R(\vex).
\eeq 
This relation plays important roles to derive equations (61) and (62). But
it does not hold for general filter functions and the simple results given in 
equations (61) and (62) are
specific to the Gaussian filter.
We summarize numerical data for the parameters$T_{GR}$ and $U_{GR}$ in  Table 3.

Generalized skewness for the gather approach is much more
complicated. If we write down  them in the form 
\beq
T_{SR}=T_{SR}+\Delta T_{SR},~~~
U_{SR}=U_{SR}+\Delta U_{SR},
\eeq
correction terms $\Delta T_{SR}$ and $\Delta U_{SR}$ are written in the manner similar to equation (39) as
follows (Appendix A.3)
\beqa
-2\Delta  T_{SR}(n)\sigma_R^2\sigma_{1R}^2&=&\frac1{6\pi^4}\int_0^\infty
{dk}\int_0^\infty{dl}\int_{-1}^1du 
\exp\lkk-\frac{(3l^2+2k^2+2klu)R^2}{2} \rkk \nn\\
& &\times k^2 l^2
P(k)P(l){(k^2+l^2+2klu)}{(k^2+l^2+klu)} R^2,\\
%%%%%%%%%%%%%%%%%%%%%%%%%
-\frac43\Delta  U_{SR}(n)\sigma_{1R}^4&=&\frac1{6\pi^4}\int_0^\infty
{dk}\int_0^\infty{dl}\int_{-1}^1du 
\exp\lkk-\frac{(3l^2+2k^2+2klu)R^2}{2} \rkk \nn\\
& &\times k^4 l^4
P(k)P(l){(k^2+l^2+klu)} (1-u^2) R^2.
\eeqa
We can calculate them explicitly as follows
\beqa
\Delta T_{SR}&=&{2^{(n+5)/2} 3^{-(n+9)/2}}\bigg[4(n+3)F\lmk
\frac{n+5}2,\frac{n+5}2,\frac52,\frac16 \rmk\nn \\
& &~~~+\frac52{(n+3)(n+5)}F\lmk
\frac{n+5}2,\frac{n+7}2,\frac52,\frac16 \rmk \nn\\
& &~~~-\frac{13(n+5)}2F\lmk\frac{n+3}2,\frac{n+7}2,\frac32,\frac16
\rmk\nn\\
& &~~~
-12(n+3)F\lmk\frac{n+5}2,\frac{n+5}2,\frac32,\frac16 \rmk\bigg],\\
%%%%%%%%%%%%%%%%%%%%%%%%%%%%%%%%%%%%
\Delta U_{SR}&=& -2^{(n+5)/2} 3^{-(n+5)/2}\bigg[4
F\lmk\frac{n+5}2,\frac{n+5}2,\frac52,\frac16 \rmk \nn \\
& &~~~+\frac{5(n+5)}9
F\lmk\frac{n+5}2,\frac{n+7}2,\frac52,\frac16 \rmk\nn \\
& &~~~-4F\lmk\frac{n+5}2,\frac{n+5}2,\frac32,\frac16 \rmk
\bigg].
\eeqa

We present numerical data of the parameters $T_{SR}$ and $U_{SR}$ in Table 4.
Note that the magnitude of
 generalized skewness $T$ and $U$ becomes very small in the
scatter approach.  
This fact becomes important in the next subsection.

\subsection{Weakly Nonlinear Genus Statistics}
In figures  3 to 5, we show the weakly nonlinear genus density smoothed
 by
 three
different  methods (fixed, gather and scatter), using  two types of
parameterizations $\nu$ and $\nu_r$. All of these curves
are smoothed by the  Gaussian filter (eq.[3]).  We plot the  normalized 
genus curves \footnote{Note that the amplitude of $G(0)$ or $G_r(0)$ are
 not changed by the 
 first-order correction of $\sigma$.} $G(\nu)/G(0)$ or  $G_r(\nu_r)/G_r(0)$ to see  deviation
 from the  symmetric
linear genus curve $\propto (1-\nu^2)\exp(-\nu^2/2)$. 
  Upper panel of figure 3  is essentially same as Fig.1 of
Matsubara (1994).

First, comparing upper and bottom panels of 
Fig.3,  we can confirm the fact experientally 
known in N-body
simulations.  Weakly nonlinear genus curves for the fixed smoothing method
 are very close to the  linear
symmetric shape in the case of $\nu_r$ parameterization ({\it e.g.}
Springel et al. 1998). For a spectral index with
$n\sim -1$, three curves for 
$\sigma=0,0.2$ and $0.4$ are nearly degenerated.

In Fig.4 we present genus curves with the gather smoothing. From Figs.3 and
4 it is apparent  that deviations from the  linear curves become larger in the
gather approach, especially in tail parts. But these  deviations become
smaller with using parameter $\nu_r$.

Bottom panel of 
Fig.4 is calculated under  conditions (gather approach and parameter
$\nu_r$) similar  to Fig.7 of Springel et al. (1998) which is
 obtained from N-body
simulations.  However overall shapes of these two are different. The minimum
value of $G_r(\nu_r)$ are  attained around the point 
$\nu_r\sim1.5$ in our result, but
this point  is $\nu_r \sim -1.5$ in theirs. 
This difference might be caused by the difference of adopted filter
function. We use the Gaussian filter but a different  kernel (a spline
kernel that is often used in SPH simulations) is adopted in their calculation
(Monaghan \& Lattanzio 1985).

In figure 5  we show results for the scatter approach. Nonlinear effects 
are more prominent than two cases analyzed earlier. As shown in
equations (46) and (56), nonlinear correction of the genus curves are
written by combination of terms proportional to parameters $S$, $T$ and
$U$. Some 
of their contribution cancel out, as realized in the case of the fixed or gather smoothing
methods. However,  amplitude of  parameters $T$ and $U$ for the scatter
approach  
becomes  very small (see
Table 4), and cancellation becomes  weaker.

Based on  the definition of the genus density (eq.[41]), nonlinear evolution
of isodensity contours is sometimes described with such terminologies as,
sponge-like (connected topology)  or meatball like (disconnected
topology).   In the case of Random Gaussian initial 
fluctuations, linear theory predicts  symmetry of the genus statistics
 with respect to the  sign 
of  density contrast $\delta$,
 and geometry  of both high and  low density tails  look 
  meat-ball like with negative genus density. 
If we use the  fixed or gather smoothing methods, nonlinear
 effects make the genus number  of a high density contour  ({\it e.g.}
 $\nu=2$) smaller, and topology of  that  region 
becomes  more  meatball-like (see figures 3 and 4). In contrast,
contour  of low density threshold ({\it e.g.}  $\nu=-2$) is transformed
in the direction of  
sponge-like topology, as quantified by the increase of genus number density (see
figures 3 and 4).   

It is not easy to understand behaviors of Fig.5 for the scatter
approach. But, changing point of views, we can discuss characters of
this approach by comparing  figures for various smoothing methods.  To
characterize nonlinear effects 
accompanied with the scatter approach, we apply the topological
interpretation mentioned 
in the last paragraph.  As shown in Fig.6 where various smoothing
methods are compared, the scatter approach makes low density regions more
meatball-like.
But high density regions are transformed in the opposite direction.
This trend shows remarkable contrast to the nonlinear gravitational effects
traced by  the simple fixed smoothing method. 

\if0%%%%%%%%%%%%%%%%%%%%%%%%%%%%%%%%%%%%%%%%%%%%%%%
 This is a remarkable difference from nonlinear
gravitational effects traced by the simple fixed smoothing method.  
   This drastic change is mainly due to large value of the
parameter $U$ in   this approach.  Around the origin $\nu\sim 0$, the
magnitude of the weakly nonlinear correction is decided by the
coefficient of $\nu$ in equation (46). They are given by (see 
eqs.[46][47])
\beq
\sigma(2.5S-4.5T+3U).
\eeq
For $\nu_r$ parameterization, we have
\beq
\sigma(2S-4.5T+3U).
\eeq
As shown in Tables  3 and 4, we  typically have $U_F\sim1.3$ and
$U_G\sim4.1$, but $U_S$ is as large 
as $\sim9$.  From these results, we can state that perturbative analysis 
of the genus statistics 
based on the multidimensional Edgeworth expansion is not suitable for
the scatter approach.
\fi%%%%%%%%%%%%%%%%%%%%%%%%%%%%%%%%%%%%%%%%%%%%%%

\subsection{Weakly Nonlinear Area Statistics}
In Figs.7 to 9, we plot the  weakly nonlinear area statistics, using equation
(50) for $\nu$ parameterization and equation (57) for $\nu_r$
parameterization. As in the analysis of the genus statistics, we
normalize amplitude of the area density by $N_3(0)$ or $N_{3r}(0)$.
Comparing upper and bottom panels of  figure 7, it is apparent that 
$\nu_r$ parameterization is very effective to keep the  original 
linear shape against
weakly nonlinear effects. This fact has been confirmed experimentally in 
N-body simulations (Ryden et al. 1989) and is quite
 similar to the situation in  the
genus statistics explained in previous subsection.  With $\nu_r$ parameterization (bottom panel of Fig.7), three
curves for $\sigma=0,0.2$ and $0.4$ are almost completely
overlapped for all spectral indexes $n$. This fact seems reasonable as we have $S_F\simeq3/2T_F$ for
the fixed smoothing method (see eq.[57] and  Tables 1,  3).

If we use the adaptive smoothing methods, weakly nonlinear correction on
$N_3(\nu)$ ($\nu$ parameterization) are  considerable  as
shown in upper panels of Figs.8 and 9. 
This correction becomes apparently smaller for the gather approach, but
not for the scatter approach.
 We have already commented
that the first-order nonlinear correction for the area statistics is
characterized by two parameters $S$ and $T$.  For the scatter approach,
$T$ parameter is very small for spectral indexes $n>-1$, and cancellation
mentioned in \S 4.3 is not effective.

\if0%%%%%%%%%%%%%%%%%%%%%%%%%%%%%%%%%%%%%%%%%
For the adaptive smoothing methods, $\nu_r$ parameterization does not
remove the nonlinear effects so clearly as in the case of
 the fixed smoothing method.
But this  parameterization is very effective,  compared with  its
performance in the genus statistics. Even for $\sigma=0.4$, deviation of 
$N_{3r}(\nu)/N_{3r}(0)$ from its linear value $e^{-\nu_r^2/2}$ (eq.[49]) 
is within $0.1$ for two adaptive methods.  For the scatter
approach  this deviation
 becomes smaller for a larger spectral
index $n$ ($-2\le n\le1$), but it very weakly depends on $n$ for the gather
approach.  
\fi%%%%%%%%%%%%%%%%%%%%%%%%%%%%%%%%%%%%%%%%

\section{Summary}
Observational analysis of galaxy clustering is one of the central issues 
in modern cosmology. Various methods have been proposed to quantify the
clustering, and many of them ({\it e.g.} topological analyses of isodensity
contour) are based on continuous smoothed density field.
 However  what we can  observe
 directly  is distribution of point-like
 galaxies.  Thus smoothing operation is crucially  important in
the  analyses of  the large-scale structure.  From theoretical point of
views, filters with spatially constant
 smoothing 
 radii are  natural choice and
have been widely   adopted so far. But we should notice that
 there are no strong
convincing reasons  
 to stick to this traditional  method.

There are  few galaxies in void regions even at  semi-nonlinear
scales. In these regions  density field obtained   with fixed smoothing radius
 might be considerably affected by the 
discreteness of (point-like)  mass elements,  and might
 hamper our analyses  of 
the cosmic structures. Adaptive smoothing is basically Lagrangian description, 
and we use nearly same number of particles to construct  smoothed density
field at each point.   Thus it is quite possible that 
 the  adaptive methods  are more efficient than the  fixed
methods  to  analyze the large-scale structure.
Actually, Springel et al. (1998) have recently 
pointed out that using adaptive filters,
signal to noise ratio of the genus statistics is improved 
even at weakly nonlinear scales $\gsim 10h^{-1}{\rm Mpc}$. 

In this article, we have developed a perturbative analysis of  adaptive 
smoothing methods  that are applied to quantify the   large-scale
structure. Even 
though adaptive methods might be  promising approaches  in   
observational cosmology,  this kind of analytic investigation
 has not been done so far.
Our targets are weakly nonlinear effects induced by  
 two typical adaptive approaches, the gather approach and the scatter
approach (Hernquist \& Katz 1989). The concept of these methods is
easily understood with 
equations (8) and (9). The gather approach is  easier to handle
analytically.  Numerical costs dealing with discrete particles' systems 
 are also  lower in this approach (Springel et al. 1998). 
Effects caused by these two  adaptive methods start from second-order
of  $\delta$ in perturbative sense.  
They   modify quantities which  characterize the 
 nonlinear mode couplings induced by gravity ({\it e.g.} Peebles 1980). 

In \S 3 we have  investigated  the skewness parameter $S$ which is a
fundamental measure to quantify asymmetry  of one point PDF.
We have shown that the  skewness for  a gather  top-hat 
filter  does not depend on the spectral index $n$ in real space, and
 very weakly depends on it ($S\simeq 35.2-0.15n$: Einstein de-Sitter
background) in redshift space. 
In the case of Gaussian filter, the skewness parameters   
show similar behaviors  both in the  scatter and gather approaches. They are 
 increasing functions  of $n$,  in contrast  to the fixed 
 smoothing method.

Next in \S 4, the genus and area statistics have been
 studied with Gaussian adaptive
filters. Our 
analysis is based on the 
multidimensional Edgeworth expansion  explored by Matsubara
(1994).
We use two quantities $\nu(\equiv\delta/\sigma)$ and $\nu_r$ to 
 parameterize isodensity contours. The latter $\nu_r$ is defined by the volume 
 fraction above  a given density threshold (Gott, Melott, \& Dickinson
 1986). It is explicitly shown that  
using  this  parameterization, two statistics with the fixed smoothing method
 are very weakly affected  by  semi-nonlinear gravitational dynamics, as
 experientally 
 confirmed by
 N-body simulations. 
For the  gather smoothing, we found that the $\nu_r$- parameterization is more 
effective to keep original linear shape of the area statistics than of the
genus statistics.

The parameters $S, T$ and $U$ 
 which characterize the nonlinear
corrections of isodensity contour depend largely on the filtering methods. 
 We can characterize nonlinear effects of these methods in somewhat
 intuitive manner, using results for the genus statistics.
The scatter approach makes  low density tails  more
 meatball-like, but high density tails are transformed in the direction
 of
  sponge-like (connected) topology.  This is a remarkable difference
 from fixed or gather smoothing methods.

\if0
 Therefore nonlinear effects for these statistics are also
 larger in adaptive  approaches. 
We have shown that perturbative analyses of genus statistics 
are not suitable for the scatter approach. In this approach, the
parameter $U$ becomes too large to treat  nonlinear effects perturbatively
even in the case of $\sigma\sim0.2$.
\fi

Our investigation in this article has been fully analytical one, using
perturbative technique of cosmological density field. Numerical analyses 
based on N-body simulations would play complementary roles to results
obtained here,  and thus are  very important. Apart from numerical
investigations, 
perturbative treatment  given in this article would be also
 developed  further
in several ways. The smoothed velocity field is crucially important
material in observational cosmology as it is supposed to be less
contaminated by effects of 
biasing (Dekel 1994, Strauss \& Willick 1995). But our observational
 information is limited to the 
line of sight peculiar velocities only at points where astrophysical objects
exist. Thus  construction of smoothed velocity field 
contains similar characters as discussed in this article ({\it e.g.}
Bernardeau \& van de Weygaert 1996).
There is another (more technical) problem that has not mentioned so far.
In this article we have only studied spherically symmetric  filter
functions. Springel et al. (1998) have shown that
signal to noise ratio of the genus curves is further improved by
using a triaxial kernel, taking   account of  tensor information
of local density field. This point must be  also
worth studying.

 \acknowledgments
The author would like to  thank J. \ Yokoyama  
for  discussion and an anonymous referee for useful comments. 
 He also thanks   H.\ Sato and N. \ Sugiyama 
for their  continuous encouragement.
This work was partially supported by the Japanese Grant
in Aid for Science Research Fund of the Ministry of Education, Science,
Sports and Culture  No.\ 3161.

\newpage

\appendix

\section{Derivations of Parameters}

In this appendix we derive expressions for the correction terms 
$\Delta S$, $\Delta T$
and $\Delta U$ given in the main text.
  First we perturbatively expand the  density contrast field smoothed by an
  adaptive filter as
\beq
\delta_A(\vex)=\delta_1(\vex)+\delta_2(\vex)+\delta_{2A}(\vex)+\cdots,
\eeq
  where $\delta_1(\vex)$ is the linear mode, $\delta_2(\vex)$ is the
  second-order mode induced by gravity, and  $\delta_{2A}$  is the
  second-order correction term caused by an  adaptive 
  smoothing (the suffix $A$ represents ``adaptive'').    Then the
  third-order moment for $\delta_A(\vex)$ is given as
\beq
\lla \delta_A^3\rra=3\lla \delta_2\delta_1^2\rra +3\lla
\delta_{2A}\delta_1^2\rra+\cdots.
\eeq
 Thus the first-order correction term   for
 the  third-order 
 moment is given as 
%\footnote{In this appendix the notation $\delta$
%   means ``difference'' not the Laplacian operator ($\nabla^2$).}
\beq
 3\lla
\delta_{2A}\delta_1^2\rra.
\eeq
In the same manner we have the following correction terms for $\lla
 \delta_A^2\nabla^2\delta_A\rra$ and  $\lla
 \nabla \delta_A\cdot\nabla\delta_a \nabla^2\delta_A\rra$ as
\beqa
& & \lla \delta_1^2\nabla^2
 \delta_{2A}\rra  +2  \lla \delta_1 \delta_{2A} \nabla^2  \delta_1\rra,\\
& &
 \lla \nabla \delta_1 \cdot \nabla \delta_1  \nabla^2 
 \delta_{2A}\rra  +2  \lla\nabla  \delta_1\nabla  \delta_{2A} \nabla^2\delta_1\rra.
 \eeqa

We can write down  the second-order correction terms $\delta_{2A}$ for the 
 gather and scatter
approaches with smoothing radius $R$ (see eqs. [10] and [11])  
\beqa
\delta_{2GR}(\vex)&=&-\frac13\delta_{1R}(\vex)R\frac{\p}{\p 
  R}\delta_{1R}(\vex)+\frac16 R\frac{d}{d 
  R}\sigma_R^2,\\
\delta_{2SR}(\vex)&=&-\frac{R}3\int  d\vex' \p_R
W(|\vex'-\vex|;R)\delta_1(\vex')\delta_{1R}(\vex').
\eeqa
where $\delta_{1R}(\vex)$ represents the smoothed linear mode,
$W(|\vex'-\vex|;R) $ is a filter function.  The variance  $\sigma_R$ of
 the matter
 fluctuations is given as 
\beqa
\sigma_{R}^2&=&\lla \delta_{1R}^2\rra+O(\delta^4)\\
&=&\int\frac{d\vek}{(2\pi)^3}w(kR)^2 P(k)+O(\delta^4),
\eeqa 
where $w(kR)$ is a Fourier transformed filter function. 
%Without
%loss of generalities,  we can discuss statistical aspects of equations
%() and () using only at point $\vex=0$.
   
For the Gaussian filter (see eq.[3]) the above equations  are  given
with the linear Fourier modes 
$\delta_1(\vek)$ as 
\beqa
\delta_{2GR}(\vex)&=&\int\frac{d\vek}{(2\pi)^3}\frac{d\vel}{(2\pi)^3}
\exp\lkk-\frac{(\vel^2+\vek^2)R^2}{2} \rkk \delta_1(\vek)
\delta_1(\vel)\frac{\vek^2R^2}3\exp[i(\vek+\vel)\cdot \vex]\nn\\
& & +\frac16 R\frac{d}{d  R}\sigma_R^2,\\
\delta_{2SR}(\vex)&=& -\int\frac{d\vek}{(2\pi)^3}\frac{d\vel}{(2\pi)^3}
\exp\lkk-\frac{(2\vel^2+\vek^2+2\vek\cdot \vel)R^2}{2} \rkk \delta_1(\vek)
\delta_1(\vel)\frac{(\vek+\vel)^2R^2}{3}\nn\\
& &\times\exp[i(\vek+\vel)\cdot \vex], 
\eeqa

Next we comment on the ensemble average of variables.
We assume that the  linear Fourier modes of density fluctuation are
random Gaussian distributed.
If variables $\{A,B,C,D\}$ obeys multivariable Gaussian distribution, we
generally 
have the  following relation
\beq
\lla ABCD\rra=\lla AB\rra \lla CD\rra + \lla AC\rra \lla BD\rra+\lla
AD\rra \lla BC\rra.
\eeq
For the linear Fourier modes the above equation becomes
\beqa
\lla\delta_1(\vek)\delta_1(\vel)\delta_1(\vem)\delta_1(\ven)\rra&=&
(2\pi)^6P(k)P(l)\delta_{Drc}(\vek+\vem)\delta_{Drc}(\vel+\ven)\nn\\
& & +(2\pi)^6P(k)P(m)\delta_{Drc}(\vek+\vel)\delta_{Drc}(\vem+\ven)\nn\\
& &+(2\pi)^6P(k)P(l)\delta_{Drc}(\vek+\ven)\delta_{Drc}(\vel+\vem).
\eeqa
Here $\delta_{Drc}(\cdot)$ is the  Dirac's delta function and $P(k)$ is
the matter power spectrum.
We evaluate expressions (A3)-(A5) using relations 
(A12)-(A13).

\subsection {Skewness}
For the gather approach the real-space representation (A6) is more
convenient. Using property (A12) we obtain  the following result
\beqa
\lla \delta_{2SR}\delta_{1R}^2\rra &=&\lla -\delta_{1R}^3 \lmk R \frac{\p}{\p
  R}\delta_{1R} \rmk \rra +\frac12 \lla \delta_{1R}^2 \rra \frac{d}{d 
  R}\sigma_R^2\\
&=&-3\sigma_{1R}^2\lla\delta_{1R}   \frac{\p}{\p
  R}\delta_{1R}\rra   +\frac12 \sigma_R^2  \frac{d}{d 
  R}\sigma_R^2\\
&=& -\sigma_R^2\frac{d}{d 
  R}\sigma_R^2.
\eeqa
The correction term for the skewness $S_G$ is written as equation (35)
\beq
\Delta S_G=\frac{\lla \delta_{2GR}\delta_{1R}^2\rra}{\sigma_R^4}
=-\frac1{\sigma_R^2}\frac{d}{d  
  R}\sigma_R^2=-\frac{d\ln \sigma_R^2}{d\ln R}.
 \eeq
For power-law models we have a simple relation $\sigma_R^2\propto
R^{-n-3}$, and  the  above
expression becomes
\beq
\Delta S_G=(n+3).
\eeq
Here we should notice that relations (A17) and (A18) do not depend on the
choice of filter functions.

Next let us evaluate the correction term for the skewness parameter in
the case of  
the  scatter approach. In this case we limit our analysis for a 
Gaussian filter (eq.[3]). From equation (A11) we have
\beqa
3\lla\delta_{2SR}(\vex)\delta_{1R}(\vex)^2\rra &=&
\int\frac{d\vek}{(2\pi)^3}\frac{d\vel}{(2\pi)^3}\frac{d\vem}{(2\pi)^3}\frac{d\ven}{(2\pi)^3}{(\vek+\vel)^2R^2}\nn\\
& &\exp\lkk-\frac{(2\vel^2+\vek^2+\vem^2+\ven^2+2\vek\cdot \vel)R^2}{2}
\rkk \nn\\
& & \times\lla  \delta_1(\vek)
\delta_1(\vel)\delta_1(\vem)\delta_1(\ven)\rra\exp[i(\vek+\vel+\vem+\ven)\cdot \vex].\nn\\  
\eeqa
Using equation (A13) we can simplify the above integral as 
\beq
2\int\frac{d\vek}{(2\pi)^3}\frac{d\vel}{(2\pi)^3}
\exp\lkk-\frac{(3\vel^2+2\vek^2+2\vek\cdot \vel)R^2}{2} \rkk
P(k)P(l){(\vek+\vel)^2} R^2.
\eeq
Note that the integrad of this expression depends only on the
information of the shape of the triangle  determined by two vectors
$\vek$ and $\vel$. 
This triangle is characterized by three quantities, namely, two sides $k=|\vek|$,
$l=|\vel|$ and cosine between them $u\equiv\vek\cdot\vel/(kl)$ with $-1\le u\le 1$. We
change variables from $\{\vek, \vel\}$ to  $\{k,l,u\}$. The volume
element is deformed  as
\beq
d\vek d\vel \Rightarrow 8\pi^2 dk dl du.
\eeq
Thus we obtain equation (39).  This equation  looks somewhat
complicated. For power-law models, however, 
we can easily  evaluate  it using {\it mathematica} (Wolfram 1996) and
finally 
obtain analytical expression (40)  given in the main text.

\subsection {Generalized Skewness for the Gather Approach}
For this approach we have the following relation for a  Gaussian filter
\beq
R \frac{\p}{\p R} \delta_{1R}(\vex)=R^2\nabla^2\delta_{1R}(\vex).
\eeq
Therefore the correction terms  (A4) and (A5) can be written by
combinations of the following  five variables \footnote{In this subsection we denote
  $\delta_{1R}(\vex)$ simply by $\delta$.}
\beq
\{\delta,~\nabla\delta,~\nabla^2\delta,~\nabla^3\delta,~\nabla^4\delta\}.
\eeq
For example, equation (A4) is written as
\beq
-\frac13 R^2\lkk \lla\delta^2\nabla^2(\delta\nabla^2\delta)\rra+2\lla
\delta^2 (\nabla^2\delta)(\nabla^2\delta) \rra-2
\lla\delta\nabla^2\delta \rra\lla\delta\nabla^2\delta  \rra \rkk
\eeq
Using property (A12), the above expression is deformed as 
\beq
-\frac13 R^2\lkk  3\lla \nabla^2\delta \nabla^2\delta\rra\lla \delta
\delta\rra+2 \lla\nabla\delta \nabla^3\delta \rra \lla \delta \delta\rra 
+3\lla \delta \nabla^4 \delta \rra 
\lla \delta \delta\rra +4\lla \delta\nabla^2 \delta \rra^2 \rkk
\eeq
The moments appeared in the above equation can be written in terms of
$P(k)$ as
\beqa
-\lla  \delta\nabla^2 \delta \rra &=&\lla\nabla \delta\nabla
\delta\rra=\int \frac{dk}{2\pi^2}k^4P(k)e^{-k^2R^2}\\
 \lla \nabla^2 \delta\nabla^2 \delta \rra &=&-\lla\nabla^3 \delta\nabla
\delta\rra= \lla \nabla^4 \delta \delta \rra=\int \frac{dk}{2\pi^2}k^6P(k)e^{-k^2R^2}
\eeqa 
For a power-law models ($P(k)\propto k^n$). These integrals  are evaluated respectively as
\beq
\sigma^2_R R^{-2}\lmk\frac{n+3}2\rmk, ~~~\sigma_R^2 R^{-4} \lmk\frac{n+3}2\rmk\lmk \frac{n+5}2\rmk. 
\eeq

With the definition  of $T$ parameter (eq.[48]) we obtain the final
result that is 
given in equation (61) as 
\beq
\Delta T_{GR}=\frac23 (n+4)
\eeq

To calculate the correction term (A5), let us use the Fourier space
representation (A10).\footnote{We  obtain the same result
  starting from equation (A6).} It is straightforward to get
\beqa
 \lla \nabla \delta_1 \cdot \nabla \delta_1  \nabla^2 
 \delta_{2A}\rra  +2  \lla\nabla  \delta_1\nabla  \delta_{2A}
 \nabla^2\delta_1\rra
 &=&\int\frac{d\vek}{(2\pi)^3}\frac{d\vel}{(2\pi)^3}\frac{d\vem}{(2\pi)^3}\frac{d\ven}{(2\pi)^3}
\frac{k^2R^2}{3}\nn\\
& &\times \exp\lkk-\frac{(\vel^2+\vek^2+\vem^2+\ven^2)R^2}{2}
\rkk \nn\\
& & \times\lla  \delta_1(\vek)
\delta_1(\vel)\delta_1(\vem)\delta_1(\ven)\rra\nn\\
& & \times\exp[i(\vek+\vel+\vem+\ven)\cdot \vex],\nn\\  
& &\times
[-(\vem\cdot\ven)(\vek+\vel)^2\nn\\
& &~~-(\vem\cdot(\vek+\vel)\ven^2)-(\ven\cdot(\vek+\vel)\vem^2) 
].
\eeqa
With equation (A13), the above integral becomes
\beqa
& &-\frac13\int\frac{d\vek}{(2\pi)^3}\frac{d\vel}{(2\pi)^3}
{k^2R^2}
 \exp\lkk-{(\vel^2+\vel^2)R^2}
\rkk   P(k) P(l) 
 [4k^2 l^2-4(\vek\cdot\vel)^2 ]\nn\\
&=&-\frac1{6\pi^4}\int_0^\infty
{dk}\int_0^\infty{dl}\int_{-1}^1du 
\exp\lkk-{(l^2+k^2)R^2} \rkk  k^6 l^4
P(k)P(l){(1-u^2)} R^2\nn\\
&=&-\frac2{9\pi^4}\int_0^\infty k^6 P(k) \exp[-k^2 R^2]\int dl l^4 P(l) \exp[-l^2 R^2].
\eeqa
For power-law models this expression becomes (see eqs.[A26]-[A28])
\beq
-\frac89 \sigma_R^4 \lmk\frac{n+3}2 \rmk^2 \lmk\frac{n+5}2  \rmk.
\eeq
Using definition of $U$ parameter (eq.[48]) we obtain equation (62) as
\beq
\Delta U_{GR}=\frac13 (n+5).
\eeq

Note that results (A29) and (A33) are not valid for general filters. Equation
(A22) that holds for the Gaussian filter plays crucial roles to derive them.

\subsection {Generalized Skewness for the Scatter Approach}
First we evaluate the correction term given in equation (A4). With the
Fourier space representation (A10) we obtain the following equation
\beqa
 \lla \delta_1^2\nabla^2
 \delta_{2A}\rra  +2  \lla \delta_1 \delta_{2A} \nabla^2
 \delta_1\rra&=&\int\frac{d\vek}{(2\pi)^3}\frac{d\vel}{(2\pi)^3}\frac{d\vem}{(2\pi)^3}\frac{d\ven}{(2\pi)^3}\frac{(\vek+\vel)^2R^2}{3}\nn\\
& &\times\exp\lkk-\frac{(2\vel^2+\vek^2+\vem^2+\ven^2+2\vek\cdot \vel)R^2}{2}
\rkk \nn\\
& & \times\lla  \delta_1(\vek)
\delta_1(\vel)\delta_1(\vem)\delta_1(\ven)\rra\exp[i(\vek+\vel+\vem+\ven)\cdot \vex],\nn\\  
& &\times [(\vek+\vel)^2+\vem^2+\ven^2]
\eeqa
With the formula (A13), this twelfth-dimensional integral becomes
\beq
\frac23\int\frac{d\vek}{(2\pi)^3}\frac{d\vel}{(2\pi)^3}
\exp\lkk-\frac{(3\vel^2+2\vek^2+2\vek\cdot \vel)R^2}{2} \rkk
P(k)P(l){(\vek+\vel)^2} R^2[(\vek+\vel)^2+\vel^2+\vek^2].
\eeq
changing variables from $d\vek d\vel$ to $dk dl du$ as shown in relation 
(A21), we obtain the result essentially same as equation   (65) as
\beqa
 \lla \delta_1^2\nabla^2
 \delta_{2A}\rra  +2  \lla \delta_1 \delta_{2A} \nabla^2
 \delta_1\rra&=&\frac1{6\pi^4}\int_0^\infty
{dk}\int_0^\infty{dl}\int_{-1}^1du 
\exp\lkk-\frac{(3l^2+2k^2+2klu)R^2}{2} \rkk \nn\\
& &\times k^2 l^2
P(k)P(l){(k^2+l^2+2klu)}{(k^2+l^2+klu)} R^2,
\eeqa
As in the case of skewness parameter, we can evaluate this complicated
integrals with {\it mathematica} and obtain equation (67).

In the same manner, equation (A5) is written as
\beqa
 \lla \nabla \delta_1 \cdot \nabla \delta_1  \nabla^2 
 \delta_{2A}\rra  +2  \lla\nabla  \delta_1\nabla  \delta_{2A}
 \nabla^2\delta_1\rra
 &=&\int\frac{d\vek}{(2\pi)^3}\frac{d\vel}{(2\pi)^3}\frac{d\vem}{(2\pi)^3}\frac{d\ven}{(2\pi)^3}
\frac{(\vek+\vel)^2R^2}{3}\nn\\
& &\times \exp\lkk-\frac{(2\vel^2+\vek^2+\vem^2+\ven^2+2\vek\cdot \vel)R^2}{2}
\rkk \nn\\
& & \times\lla  \delta_1(\vek)
\delta_1(\vel)\delta_1(\vem)\delta_1(\ven)\rra\nn\\
& &\times\exp[i(\vek+\vel+\vem+\ven)\cdot \vex],\nn\\  
& &\times
[-(\vem\cdot\ven)(\vek+\vel)^2\nn\\
& &~~-(\vem\cdot(\vek+\vel)\ven^2)-(\ven\cdot(\vek+\vel)\vem^2) 
].
\eeqa
This expression is simplified to the following form
\beqa
& &\frac13\int\frac{d\vek}{(2\pi)^3}\frac{d\vel}{(2\pi)^3}
{(\vek+\vel)^2R^2}
 \exp\lkk-\frac{(2\vel^2+\vek^2+\vem^2+\ven^2+2\vek\cdot \vel)R^2}{2}
\rkk \nn\\
& & \times P(k) P(l) 
 [4k^2 l^2-4(\vek\cdot\vel)^2 ].
\eeqa
Again, changing variables, we obtain the expression (66) as
\beqa
\lla \nabla \delta_1 \cdot \nabla \delta_1  \nabla^2 
 \delta_{2A}\rra  +2  \lla\nabla  \delta_1\nabla  \delta_{2A}
 \nabla^2\delta_1\rra
 &=&\frac1{6\pi^4}\int_0^\infty
{dk}\int_0^\infty{dl}\int_{-1}^1du {(1-u^2)} R^2\nn\\
& &\times\exp\lkk-\frac{(3l^2+2k^2+2klu)R^2}{2} \rkk \nn\\
& &\times k^4 l^4
P(k)P(l){(k^2+l^2+2klu)}.
\eeqa
For power-law
models  we can   evaluate this integral using {\it mathematica} and
obtain   equation (68).

\newpage
%\begin{references}

\newpage

\begin{center}
TABLE 1\\
{\sc skewness  for the
  gather approach (Gaussian filter)}\\ 
\ \\
\begin{tabular}{ccccccc}
\hline\hline
 spectral index $n$  &1& 0 &-1& -2& -3  \\
\hline
 $S_F(n)$  ~~ &~~ 3.029~~ &~~ 3.144~~ &~~3.468 ~~ &~~4.022
 ~~&4.857 \\
 $\Delta S_G(n)$  ~~ &~~ 4.000~~ &~~ 3.000~~ &~~2.000 ~~ &~~1.000
 ~~&0 \\
 $S_G(n)$  ~~ &~~ 7.029~~ &~~ 6.144~~ &~~5.468 ~~ &~~5.022 ~~&4.857 \\
\hline
\end{tabular}
\end{center}

\begin{center}
TABLE 2\\
{\sc skewness  for the
  scatter approach}\\ 
\ \\
\begin{tabular}{ccccccc}
\hline\hline
 spectral index $n$  &1& 0 &-1& -2& -3  \\
\hline
 $S_F(n)$  ~~ &~~ 3.029~~ &~~ 3.144~~ &~~3.468 ~~ &~~4.022
 ~~&4.857 \\
 $\Delta_S S(n)$  ~~ &~~ 3.031~~ &~~ 2.576~~ &~~2.045 ~~ &~~1.277
 ~~&0 \\
 $S_S(n)$  ~~ &~~ 6.060~~ &~~ 5.720~~ &~~5.513 ~~ &~~5.299 ~~&4.857 \\
\hline
\end{tabular}
\end{center}

%\newpage

\begin{center}
TABLE 3\\
{\sc generalized skewness  for the
  gather approach}\\ 
\ \\
\begin{tabular}{ccccccc}
\hline\hline
 spectral index $n$  &1& 0 &-1& -2& -3  \\
\hline
 $T_F(n)$  ~~ &~~ 2.020~~ &~~ 2.096~~ &~~2.312 ~~ &~~2.681
 ~~&3.238 \\
 $\Delta T_G(n)$  ~~ &~~ 3.333~~ &~~ 2.667~~ &~~2.000 ~~ &~~1.333
 ~~&0.667 \\
 $T_G(n)$  ~~ &~~ 5.353~~ &~~ 4.763~~ &~~4.312 ~~ &~~4.014 ~~&3.905 \\
\hline
 $U_G(n)$  ~~ &~~ 1.431~~ &~~ 1.292~~ &~~1.227 ~~ &~~1.222
 ~~&1.272 \\
 $\Delta U_G(n)$  ~~ &~~ 2.000~~ &~~ 1.667~~ &~~1.333 ~~ &~~1.000
 ~~&0.667 \\
 $U_G(n)$  ~~ &~~ 3.431~~ &~~ 2.959~~ &~~2.560 ~~ &~~2.222 ~~&1.929 \\
\hline
\end{tabular}
\end{center}

\begin{center}
TABLE 4\\
{\sc generalized skewness  for the
  scatter approach}\\ 
\ \\
\begin{tabular}{ccccccc}
\hline\hline
 spectral index $n$  &1& 0 &-1& -2& -3  \\
\hline
 $T_F(n)$  ~~ &~~ 2.020~~ &~~ 2.096~~ &~~2.312 ~~ &~~2.681
 ~~&3.238 \\
 $\Delta T_S(n)$  ~~ &~~ -2.082~~ &~~ -1.908~~ &~~-1.723 ~~ &~~-1.451
 ~~&-0.963 \\
 $T_S(n)$  ~~ &~~ -0.0623~~ &~~ 0.1882~~ &~~0.5892 ~~ &~~1.230 ~~&2.275 \\
\hline
 $U_F(n)$  ~~ &~~ 1.431~~ &~~ 1.292~~ &~~1.227 ~~ &~~1.222
 ~~&1.272 \\
 $\Delta U_S(n)$  ~~ &~~ -1.265~~ &~~ -1.145~~ &~~-1.027 ~~ &~~-0.8916
 ~~&-0.7105 \\
 $U_S(n)$  ~~ &~~ 0.1662~~ &~~ 0.1474~~ &~~0.2000 ~~&~~0.3301~~&0.5611\\
\hline
\end{tabular}
\end{center}

%\if0%%%%%%%%%%%%%%%%%%%%%%%%%%%%%%%%%%%%%%%%%%%%%%%%%%%%%%%%%%%%%%%

\begin{figure}[h]
 \begin{center}
 \epsfxsize=15.2cm
 \begin{minipage}{\epsfxsize} \epsffile{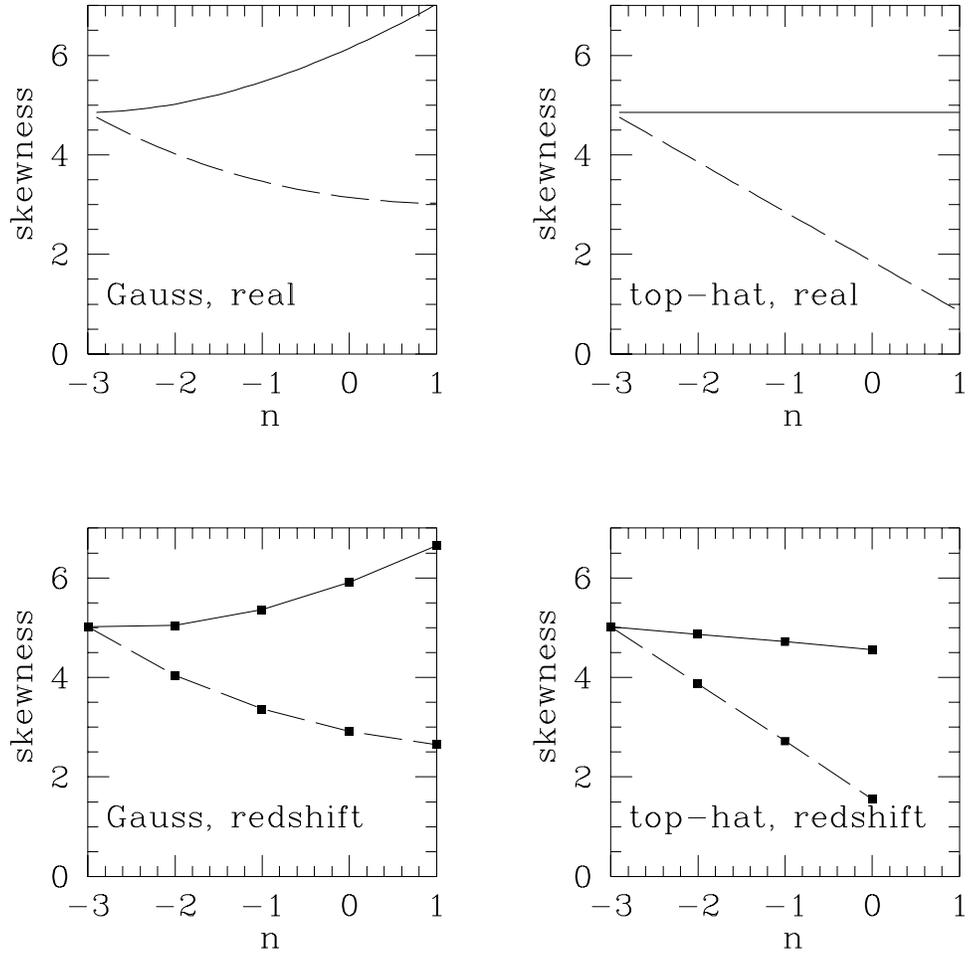} \end{minipage}
 \end{center}
\caption[]{Skewness for the gather approach  in  real  and
   redshift spaces.  We use two kinds of filters (Gaussian
 and top-hat filters).
 The dashed-lines represent 
  results for the traditional 
fixed smoothing method and the solid lines for the gather approach.
Numerical data in the  redshift space are based on Table 2 and 3 of Hivon et
al. (1995).  }
\end{figure}

\begin{figure}[h]
 \begin{center}
 \epsfxsize=15.2cm
 \begin{minipage}{\epsfxsize} \epsffile{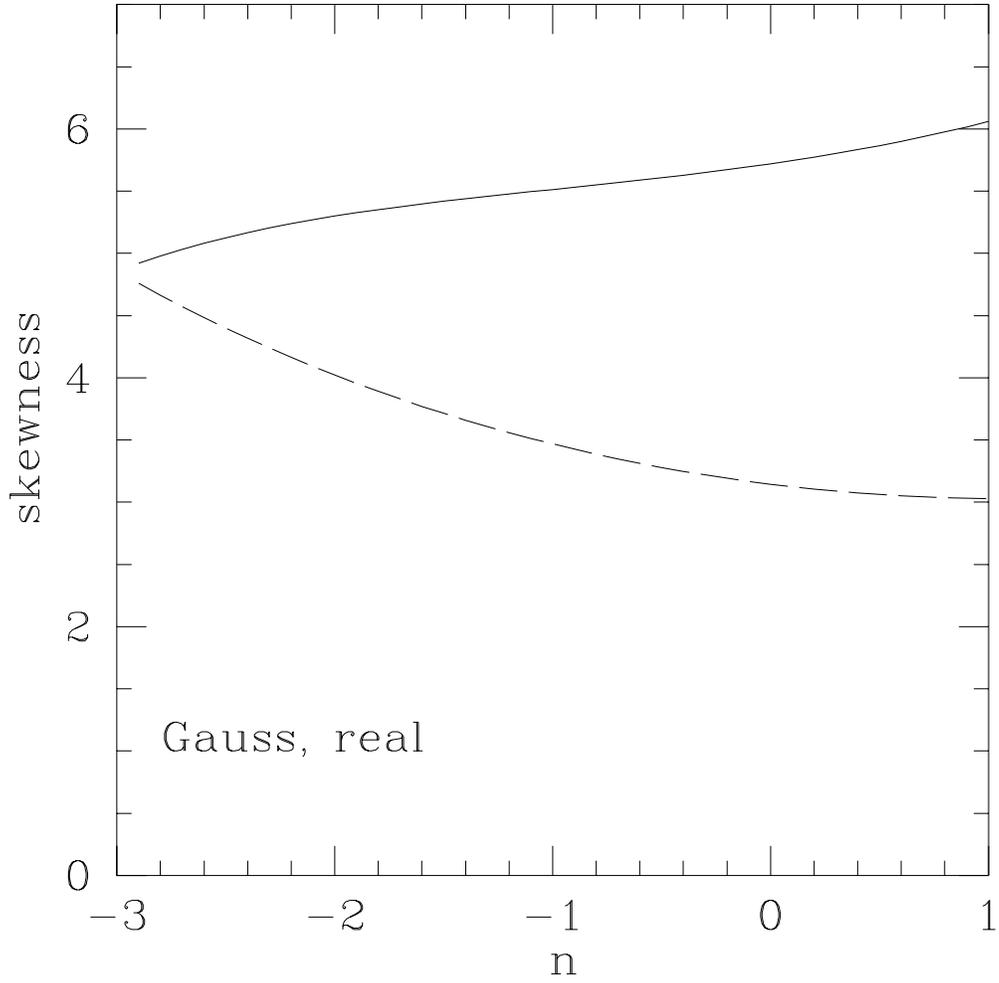} \end{minipage}
 \end{center}
\caption[]{Real space skewness for the scatter approach  with the 
  Gaussian filters. The dashed-line represents results for  
  the fixed smoothing method and the solid line for the scatter approach.}
\end{figure}

\begin{figure}[h]
 \begin{center}
 \epsfxsize=15.2cm
 \begin{minipage}{\epsfxsize} \epsffile{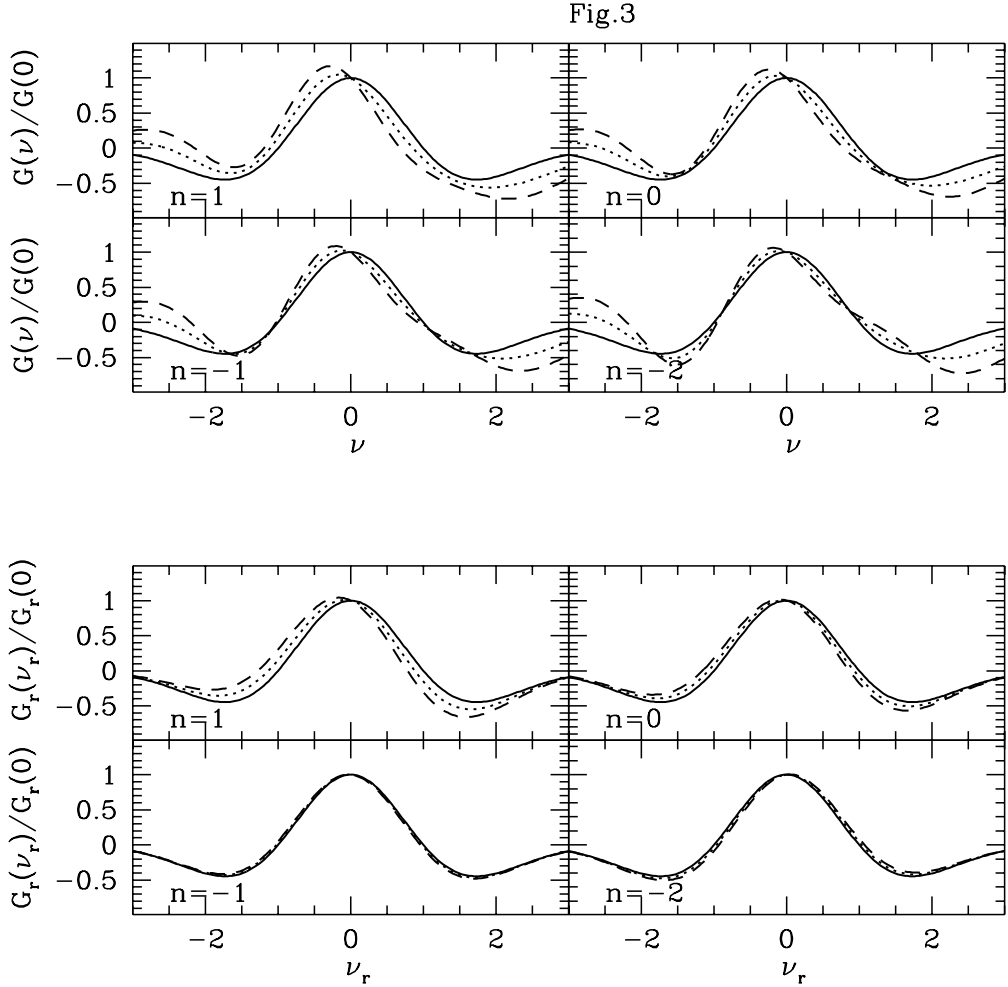} \end{minipage}
 \end{center}
\caption[]{Normalized genus density for the fixed Gaussian smoothing (see
  Matsubara 1994). Upper panel corresponds to 
  $\nu$- parameterization and lower to  $\nu_r$- parameterization. The solid curves represent linear curves
  (eq.[45]). Dotted-lines, dashed-lines show $\sigma=0.2$, and
  $\sigma=0.4$ respectively.}
\end{figure}

\begin{figure}[h]
 \begin{center}
 \epsfxsize=15.2cm
 \begin{minipage}{\epsfxsize} \epsffile{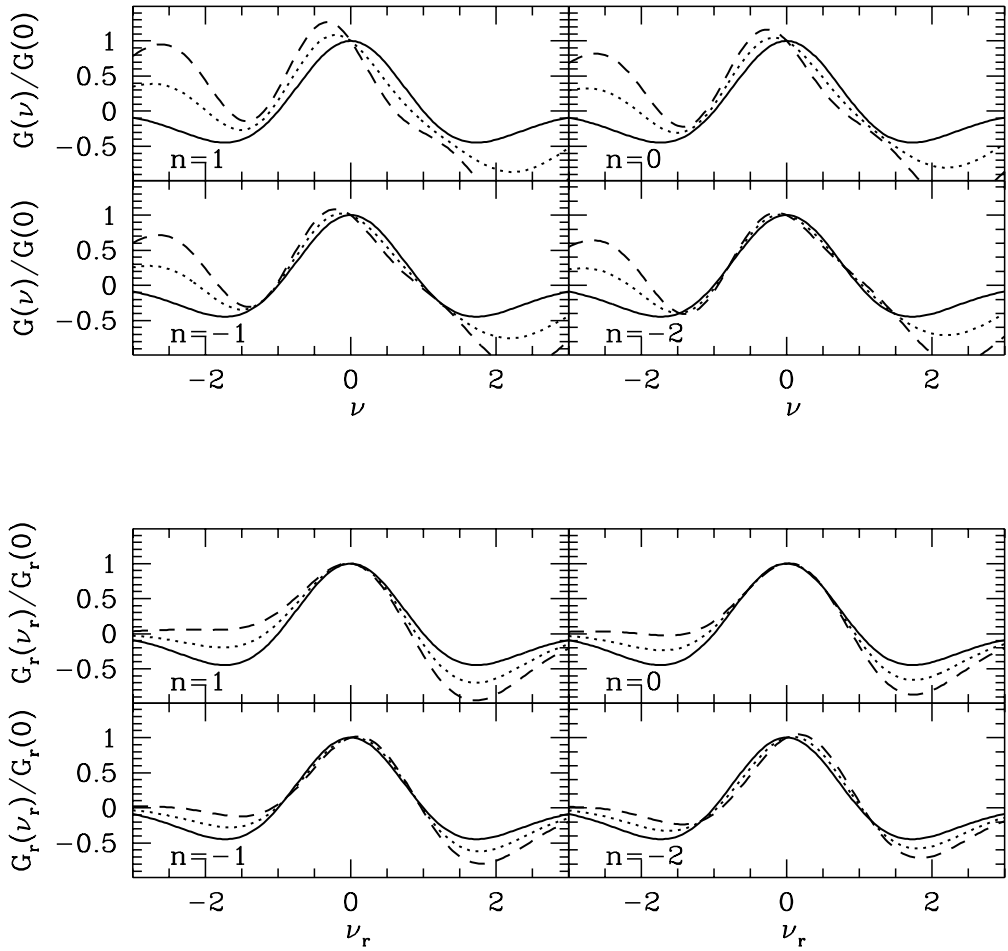} \end{minipage}
 \end{center}
\caption[]{Same as Fig.3 but with the gather approach. }
\end{figure}

\begin{figure}[h]
 \begin{center}
 \epsfxsize=15.2cm
 \begin{minipage}{\epsfxsize} \epsffile{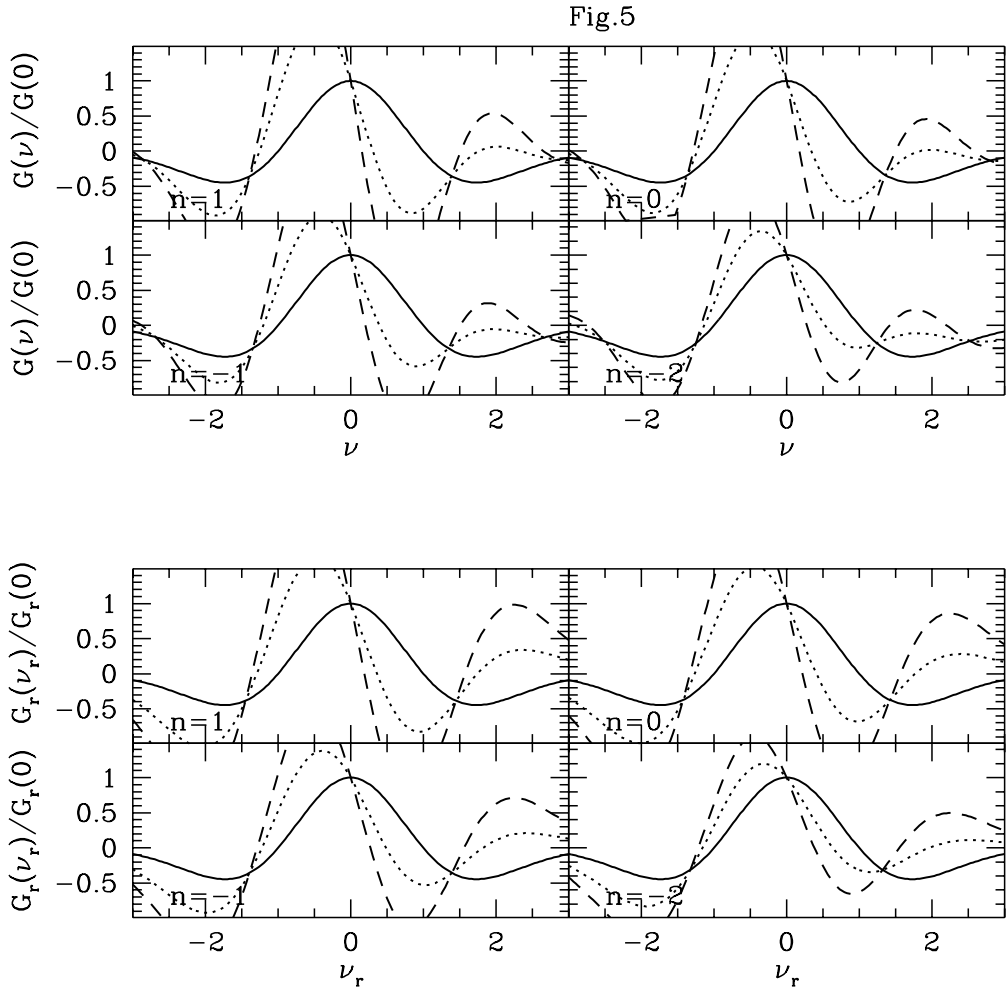} \end{minipage}
 \end{center}
\caption[]{Same as Fig.3 but with  the scatter approach. }
\end{figure}

\begin{figure}[h]
 \begin{center}
 \epsfxsize=15.2cm
 \begin{minipage}{\epsfxsize} \epsffile{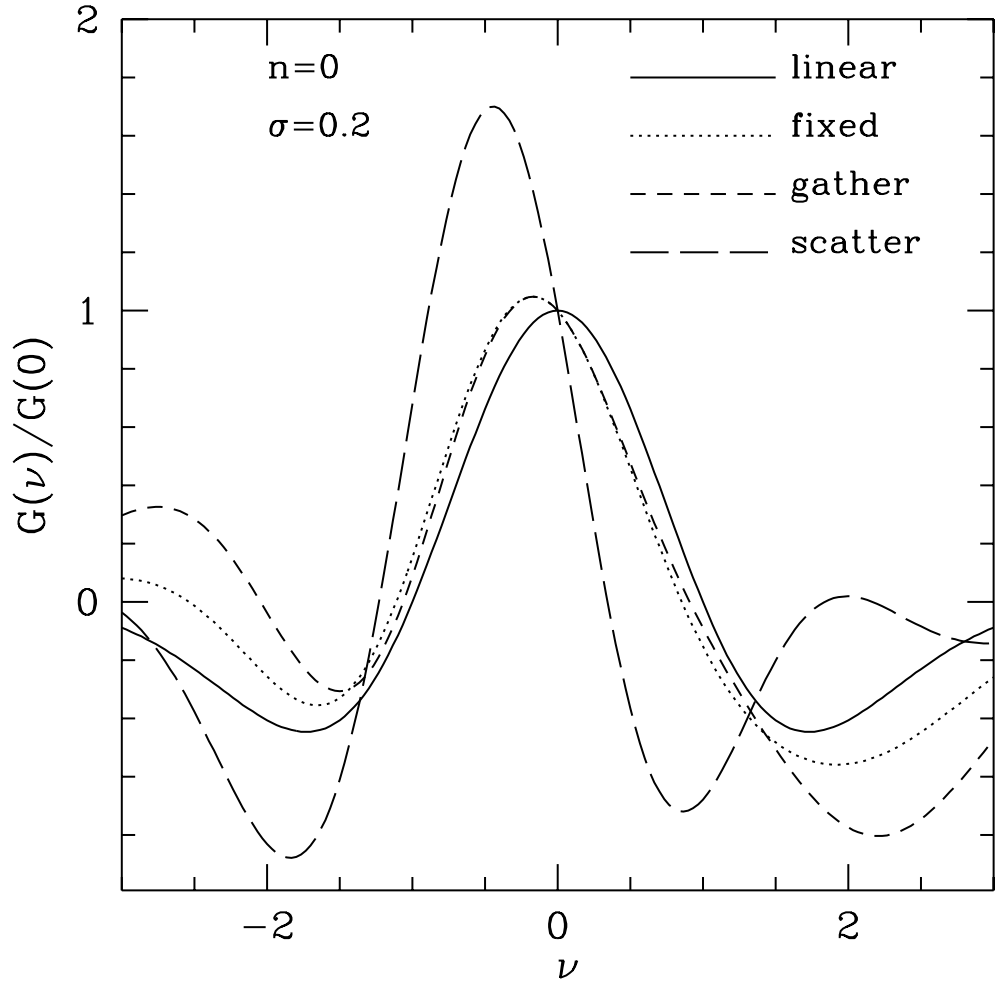} \end{minipage}
 \end{center}
\caption[]{Nonlinear Genus curves for various smoothing methods. The
  solid line corresponds to the linear analysis, dotted to the fixed smoothing,
  short-dased to the gather approach, and long-dased to the scatter
  approach. We fix the spectral index at $n=0$ and nonlinearity at $\sigma=0.2$. }
\end{figure}

\begin{figure}[h]
 \begin{center}
 \epsfxsize=15.2cm
 \begin{minipage}{\epsfxsize} \epsffile{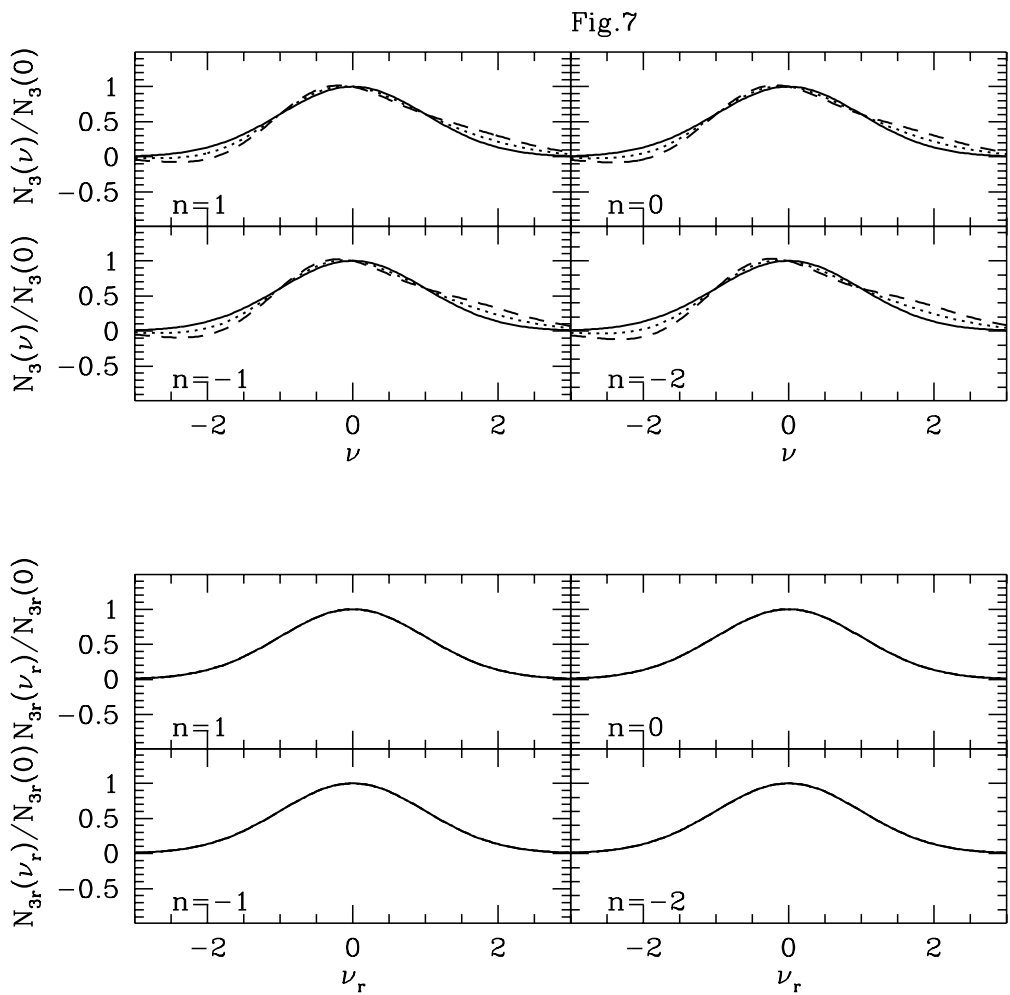} \end{minipage}
 \end{center}
\caption[]{Normalized area statistics  for the fixed Gaussian smoothing. Upper panel corresponds to 
  $\nu$ parameterization and lower to  $\nu_r$ parameterization. The
  solid curves represent the linear prediction 
  (eq.[45]). Dotted-lines, dashed-lines show $\sigma=0.2$, and
  $\sigma=0.4$ respectively. }
\end{figure}

\begin{figure}[h]
 \begin{center}
 \epsfxsize=15.2cm
 \begin{minipage}{\epsfxsize} \epsffile{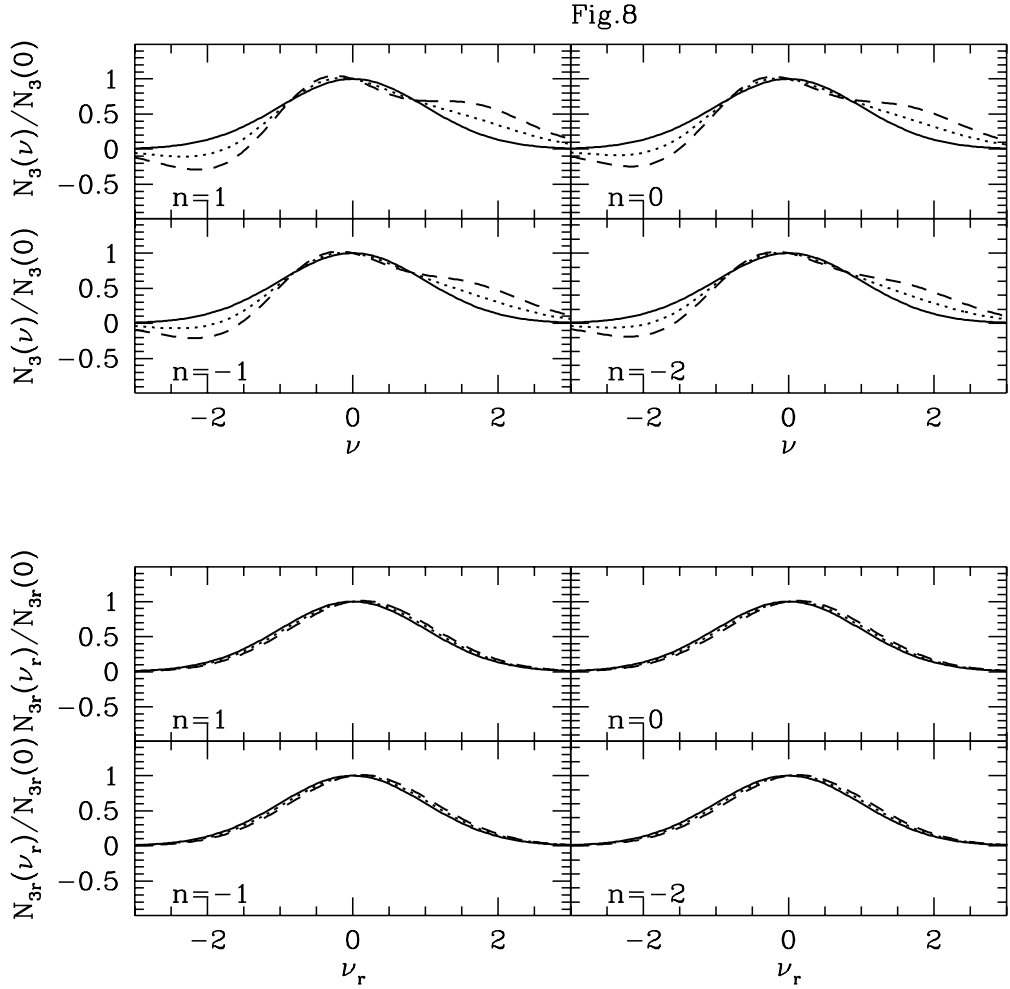} \end{minipage}
 \end{center}
\caption[]{Same as Fig.6 but for the  gather approach. }
\end{figure}

\begin{figure}[h]
 \begin{center}
 \epsfxsize=15.2cm
 \begin{minipage}{\epsfxsize} \epsffile{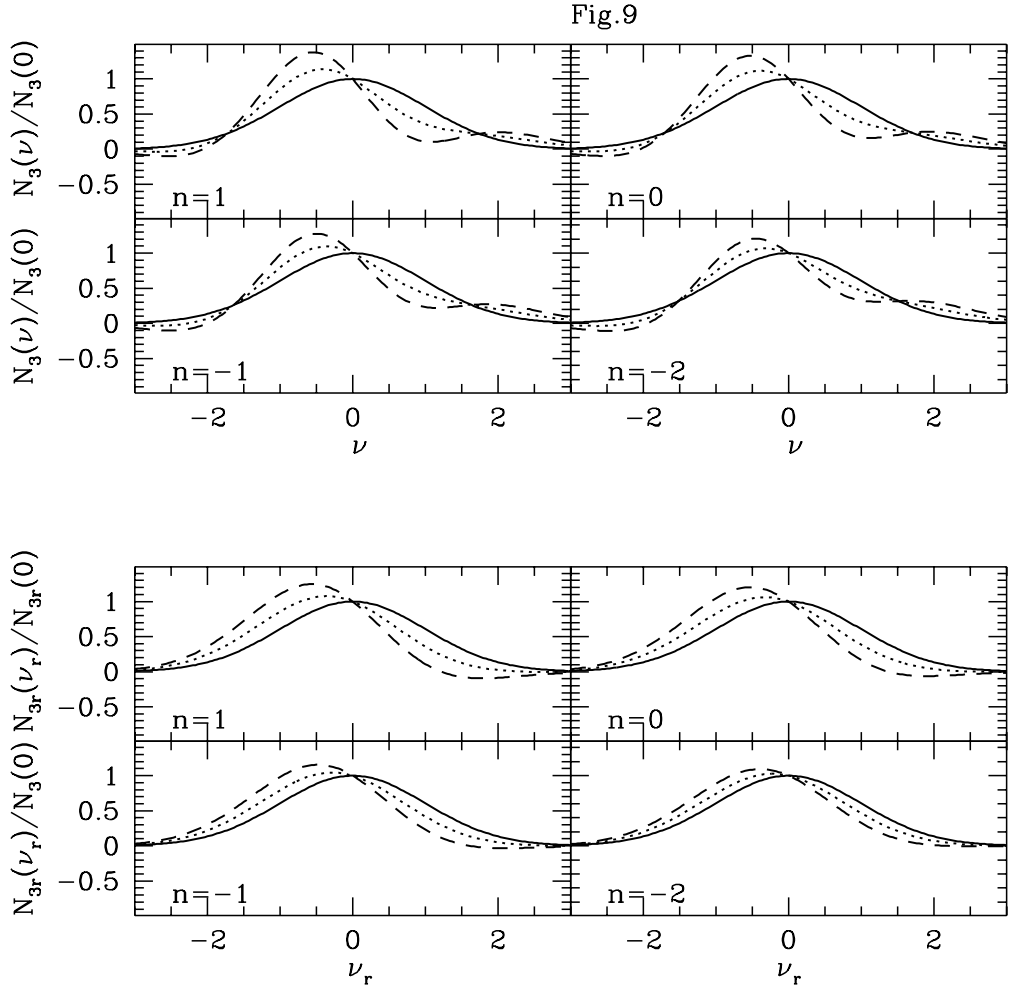} \end{minipage}
 \end{center}
\caption[]{Same as Fig.6 but for the  scatter approach. }
\end{figure}
\end{document}